\documentclass[zpreprint,zbstnp]{zeus_paper}
\usepackage[english]{babel}
\newcommand{\Zdetdesc}{%
A detailed description of the ZEUS detector can be found 
elsewhere~\cite{zeus:1993:bluebook}. A brief outline of the 
components that were most relevant for this analysis is given
below.\xspace}

\chardef\usc=95
\chardef\til=126
\catcode`\@=11 
\DeclareRobustCommand\xdotspace{\futurelet\@let@token\@xdotspace}
\def\@xdotspace{%
  \ifx\@let@token.\else
  \ifx\@let@token\bgroup.\else
  \ifx\@let@token\egroup.\else
  \ifx\@let@token\/.\else
  \ifx\@let@token\ .\else
  \ifx\@let@token~.\else
  \ifx\@let@token!.\else
  \ifx\@let@token,.\else
  \ifx\@let@token:.\else
  \ifx\@let@token;.\else
  \ifx\@let@token?.\else
  \ifx\@let@token/.\else
  \ifx\@let@token'.\else
  \ifx\@let@token).\else
  \ifx\@let@token-.\else
  \ifx\@let@token\@xobeysp.\else
  \ifx\@let@token\space.\else
  \ifx\@let@token\@sptoken.\else
   .\space
   \fi\fi\fi\fi\fi\fi\fi\fi\fi\fi\fi\fi\fi\fi\fi\fi\fi\fi}
\catcode`\@=12 

\newcommand{\stru}[2]{%
   \relax\ifmmode\hbox{\vrule height#1 depth#2 width0pt}%
   \else\vrule height#1 depth#2 width0pt\fi}

\newcommand{\Ronum}[1]{\uppercase\expandafter{\romannumeral#1}}
\newcommand{\ronum}[1]{\expandafter{\romannumeral#1}}
\DeclareRobustCommand{\LaTeXZ}{%
  \LaTeX\kern-.05em4\kern-.1em
  {\raisebox{-0.2ex}{$\scriptstyle\text{ZEUS}$}}\xspace}

\DeclareMathAlphabet{\mathbf}{OT1}{cmr}{bx}{sl}
\newcommand{\eVdist}{\kern-0.06667em}

\newcommand{\Gev}{{\text{Ge}\eVdist\text{V\/}}}

\newcommand{\gev}{{\,\text{Ge}\eVdist\text{V\/}}}

\newcommand{\pb}{\,\text{pb}}

\newcommand{\Tesla}{\,\text{T}}

\newcommand{\slashfrac}[2]{%
  \raisebox{0.5ex}{\ensuremath #1}\kern-0.12em/\kern-0.08em
  \raisebox{-.8ex}{\ensuremath #2}}

\newcommand{\sqr}[3]{%
    {\vcenter{\hrule height.#3ex\hbox{\vrule width.#2ex height#1ex
     \kern#1ex\vrule width.#3ex}\hrule height.#2ex}}}

\catcode`\@=11 
\newcommand{\parenbar}{\mathpalette\p@renb@r}
\def\p@renb@r#1#2{\vbox{%
  \ifx#1\scriptscriptstyle \dimen@.7em\dimen@ii.2em\else
  \ifx#1\scriptstyle \dimen@.8em\dimen@ii.25em\else
  \dimen@1em\dimen@ii.4em\fi\fi \offinterlineskip
  \ialign{\hfill##\hfill\cr
    \vbox{\hrule width\dimen@ii}\cr
    \noalign{\vskip-.3ex}%
    \hbox to\dimen@{$\mathchar300\hfil\mathchar301$}\cr
    \noalign{\vskip-.3ex}%
    $#1#2$\cr}}}
\catcode`\@=12 

\newcommand{\IP}{{\rm I$\kern-0.01667em$P}\xspace}

\mathchardef\qsm=63
\mathchardef\pls=43
\mathchardef\mns=512
\mathchardef\plm=518
\mathchardef\eql=61
\mathchardef\smallleft=300
\mathchardef\smallright=301
\mathchardef\les=316
\mathchardef\gre=318
\mathchardef\leq=532
\mathchardef\grq=533

\catcode`\@=11 
\newcounter{pict@width}
\newcounter{pict@height}
\newlength{\pict@scale}
\newcommand{\psfigadd}[4]{%
\setcounter{pict@width}{1*\ratio{#2+\pict@scale/2}{\pict@scale}}
\setcounter{pict@height}{1*\ratio{#3+\pict@scale/2}{\pict@scale}}
\setlength{\unitlength}{\pict@scale}
\hbox to #2{\hspace{-\fill}\begin{picture}(\thepict@width,\thepict@height)
\put(0,0){\psfig{figure=#1,width=#2,height=#3,clip=}}
\SetScale{0.283466457}
\SetWidth{1.763889}
{#4}
\end{picture}}
}
\newcounter{pict@widthfst}
\newcounter{pict@widthscd}
\newcounter{pict@widthtot}
\newcommand{\psfigaddtwo}[7]{%
\setcounter{pict@widthfst}{1*\ratio{#2+\pict@scale/2}{\pict@scale}}
\setcounter{pict@widthscd}{1*\ratio{#2+#4+\pict@scale/2}{\pict@scale}}
\setcounter{pict@widthtot}{1*\ratio{#2+#4+#6+\pict@scale/2}{\pict@scale}}
\setcounter{pict@height}{1*\ratio{#3+\pict@scale/2}{\pict@scale}}
\setlength{\unitlength}{\pict@scale}
\hbox{\hspace{-\fill}\begin{picture}(\thepict@widthtot,\thepict@height)
\put(0,0){\psfig{figure=#1,width=#2,height=#3,clip=}}
\put(\thepict@widthscd,0){\psfig{figure=#5,width=#6,height=#3,clip=}}
\SetScale{0.283466457}
\SetWidth{1.763889}
{#7}
\end{picture}}
}
\newcommand{\psfigror}[4]{%
\setcounter{pict@width}{1*\ratio{#2+\pict@scale/2}{\pict@scale}}
\setcounter{pict@height}{1*\ratio{#3+\pict@scale/2}{\pict@scale}}
\setlength{\unitlength}{\pict@scale}
\hbox{\begin{picture}(\thepict@width,\thepict@height)
\put(0,\thepict@height){\psfig{figure=#1,width=#3,height=#2,clip=,angle=270}}
\SetScale{0.283466457}
\SetWidth{1.763889}
{#4}
\end{picture}}
}
\newcommand{\psfigrol}[4]{%
\setcounter{pict@width}{1*\ratio{#2+\pict@scale/2}{\pict@scale}}
\setcounter{pict@height}{1*\ratio{#3+\pict@scale/2}{\pict@scale}}
\setlength{\unitlength}{\pict@scale}
\hbox{\begin{picture}(\thepict@width,\thepict@height)
\put(0,0){\psfig{figure=#1,width=#3,height=#2,clip=,angle=90}}
\SetScale{0.283466457}
\SetWidth{1.763889}
{#4}
\end{picture}}
}
\catcode`\@=12 
\newlength\listtextwidth

\catcode`\@=11 
\newlength{\@tabfninsert}
\newlength{\@tabfnwidth}
\newcommand{\tabfootnote}[2]{%
  \setlength{\@tabfninsert}{0.8em}
  \setlength{\@tabfnwidth}{\textwidth}
  \addtolength{\@tabfnwidth}{-\@tabfninsert}
  \addtolength{\@tabfnwidth}{-0.4em}
  \noindent\makebox[\@tabfninsert][r]{\footnotesize$^{#1}$\hfil}\hfill%
  \parbox[t]{\@tabfnwidth}{\footnotesize #2\hfill}}
\catcode`\@=12 

\def\citeCTD{{\cite{%
nim:a279:290,*npps:b32:181,*nim:a338:254%
}}\xspace}
\def\citeMVD{{\cite{%
nim:a581:656%
}}\xspace}
\def\citeCAL{{\cite{%
nim:a309:77,*nim:a309:101,*nim:a321:356,*nim:a336:23%
}}\xspace}

\begin{document}

\prepnum{{DESY--09--56}}

\title{
Measurement of charm and beauty production
in deep inelastic $ep$ scattering
from decays into muons at HERA 
}                                                       
                    
\author{ZEUS Collaboration}
\date{April 2009}

\abstract{The production of charm and beauty quarks
in $ep$ interactions  has been measured  with the ZEUS detector at HERA
for squared four-momentum exchange $Q^2>20\gev^2$, using an
integrated luminosity of $126$~pb$^{-1}$.
Charm and beauty quarks were identified through
their decays into muons.
Differential cross sections were measured for muon
transverse momenta $p_T^{\mu}>1.5\gev$ and pseudorapidities
 $-1.6<\eta^{\mu}<2.3$,
as a function of $p_T^{\mu}$, $\eta^{\mu}$, $Q^2$ and Bjorken $x$.
The charm and beauty contributions to the proton structure function $F_2$
were also extracted. The results agree with previous
measurements based on independent techniques and are well described by 
QCD predictions.}

\makezeustitle

\def\3{\ss}                                                                                        
\newcommand{\address}{ }                                                                           
\pagenumbering{Roman}                                                                              
                                                   %
\begin{center}                                                                                     
{                      \Large  The ZEUS Collaboration              }                               
\end{center}                                                                                       
  S.~Chekanov,                                                                                     
  M.~Derrick,                                                                                      
  S.~Magill,                                                                                       
  B.~Musgrave,                                                                                     
  D.~Nicholass$^{   1}$,                                                                           
  \mbox{J.~Repond},                                                                                
  R.~Yoshida\\                                                                                     
 {\it Argonne National Laboratory, Argonne, Illinois 60439-4815, USA}~$^{n}$                       
\par \filbreak                                                                                     
  M.C.K.~Mattingly \\                                                                              
 {\it Andrews University, Berrien Springs, Michigan 49104-0380, USA}                               
\par \filbreak                                                                                     
  P.~Antonioli,                                                                                    
  G.~Bari,                                                                                         
  L.~Bellagamba,                                                                                   
  D.~Boscherini,                                                                                   
  A.~Bruni,                                                                                        
  G.~Bruni,                                                                                        
  F.~Cindolo,                                                                                      
  M.~Corradi,                                                                                      
\mbox{G.~Iacobucci},                                                                               
  A.~Margotti,                                                                                     
  R.~Nania,                                                                                        
  A.~Polini\\                                                                                      
  {\it INFN Bologna, Bologna, Italy}~$^{e}$
% ...
\newpage                                                                                           
\enlargethispage{5cm}                                                                              
$^{\    1}$ also affiliated with University College London,                                        
United Kingdom\\                                                                                   
% ...
$^{\    5}$ supported by the research grant No. 1 P03B 04529 (2005-2008) \\                        
% ...
%
\newpage   
                                                           %
                                                           %
\begin{tabular}[h]{rp{14cm}}                                                                       
$^{a}$ &  supported by the Natural Sciences and Engineering Research Council of Canada (NSERC) \\  
$^{b}$ &  supported by the German Federal Ministry for Education and Research (BMBF), under        
          contract Nos. 05 HZ6PDA, 05 HZ6GUA, 05 HZ6VFA and 05 HZ4KHA\\                            
$^{c}$ &  supported in part by the MINERVA Gesellschaft f\"ur Forschung GmbH, the Israel Science   
          Foundation (grant No. 293/02-11.2) and the US-Israel Binational Science Foundation \\    
$^{d}$ &  supported by the Israel Science Foundation\\                                             
$^{e}$ &  supported by the Italian National Institute for Nuclear Physics (INFN) \\                
$^{f}$ &  supported by the Japanese Ministry of Education, Culture, Sports, Science and Technology 
          (MEXT) and its grants for Scientific Research\\                                          
$^{g}$ &  supported by the Korean Ministry of Education and Korea Science and Engineering          
          Foundation\\                                                                             
$^{h}$ &  supported by the Netherlands Foundation for Research on Matter (FOM)\\                   
$^{i}$ &  supported by the Polish State Committee for Scientific Research, project No.             
          DESY/256/2006 - 154/DES/2006/03\\                                                        
$^{j}$ &  partially supported by the German Federal Ministry for Education and Research (BMBF)\\   
$^{k}$ &  supported by RF Presidential grant N 1456.2008.2 for the leading                         
          scientific schools and by the Russian Ministry of Education and Science through its      
          grant for Scientific Research on High Energy Physics\\                                   
$^{l}$ &  supported by the Spanish Ministry of Education and Science through funds provided by     
          CICYT\\                                                                                  
$^{m}$ &  supported by the Science and Technology Facilities Council, UK\\                         
$^{n}$ &  supported by the US Department of Energy\\                                               
$^{o}$ &  supported by the US National Science Foundation. Any opinion,                            
findings and conclusions or recommendations expressed in this material                             
are those of the authors and do not necessarily reflect the views of the                           
National Science Foundation.\\                                                                     
$^{p}$ &  supported by the Polish Ministry of Science and Higher Education                         
as a scientific project (2006-2008)\\                                                              
$^{q}$ &  supported by FNRS and its associated funds (IISN and FRIA) and by an Inter-University    
          Attraction Poles Programme subsidised by the Belgian Federal Science Policy Office\\     
$^{r}$ &  supported by an FRGS grant from the Malaysian government\\                               
\end{tabular}                                                                                      
                                                           %
                                                           %

\pagenumbering{arabic} 
\pagestyle{plain}
\section{Introduction}
\label{sec-int}

The measurement of charm and beauty production in deep inelastic scattering
(DIS) provides a stringent test of quantum chromodynamics (QCD) since the
large quark masses provide hard scales that make
perturbative calculations applicable.
At leading order, heavy quarks (HQs) are produced in DIS  via boson--gluon
fusion (BGF) ($\gamma^* g \rightarrow q \bar{q}$).
A precise measurement of HQ production in DIS therefore
provides a direct constraint on the gluon parton density function (PDF)
of the proton.

Charm production in DIS at HERA has been measured previously using  
reconstructed charmed mesons~\cite{pl:b407:402,*epj:c12:35,*pl:b528:199,*epj:c38:447,pr:d69:012004} or inclusively by
exploiting the long lifetime of charmed hadrons~\cite{epj:c40:349,*epj:c45:23}.
Beauty production in DIS has been studied in events with muons and
jets~\cite{pl:b599:173,epj:c41:453} and from 
lifetime information~\cite{epj:c40:349,*epj:c45:23}.
The existing data are generally in good agreement with next-to-leading-order
(NLO) QCD predictions.
The largest differences were observed
in the muon analyses~\cite{pl:b599:173,epj:c41:453} 
where the measured beauty cross section was about two standard
deviations above the theoretical expectation.

In this paper, a simultaneous measurement of beauty and charm production
using semi-leptonic (SL) decays into muons is presented.
The fractions of muons originating from charm,
beauty and light flavours (LF) were extracted by exploiting three
discriminating variables: the muon impact parameter,
the muon momentum component transverse to the associated jet axis and
the missing transverse momentum, which is sensitive to the neutrino from SL 
decays.

The analysis focused on data with large squared four-momentum exchange
at the electron vertex, $Q^2$, where 
charm measurements based on muons are competitive with those based
on identified charmed mesons.

The cross sections for muons from charm and beauty decays
were measured for $Q^2>20\gev^2$,
muon transverse momenta $p_T^{\mu}>1.5\gev$ and
pseudorapidities\footnote{The ZEUS coordinate system is a right-handed Cartesian system, with the $Z$
axis pointing in the proton beam direction, referred to as the ``forward
direction'', and the $X$ axis pointing towards the centre of HERA.
The pseudorapidity is defined as $\eta=-\ln\left(\tan\frac{\theta}{2}\right)$,
where $\theta$ is the polar angle.
\xspace} $-1.6<\eta^{\mu}<2.3$ 
as a function of $p_T^{\mu}$, $\eta^{\mu}$, $Q^2$, and
of the Bjorken scaling variable $x$~\cite{pr:179:1547}
and compared to QCD predictions. 
The muon cross sections, measured in bins of $x$ and $Q^2$, were
used to extract the heavy quark contributions to the proton structure 
function $F_2$ which were compared to previous results and to QCD predictions.

The data used in this analysis were collected with the ZEUS
detector in the 2005 running period during which HERA collided
electrons with energy $E_e=27.5\gev$ 
with protons with $E_p=920\gev$ corresponding to
a centre-of-mass energy $\sqrt{s} = 318\gev$. The corresponding integrated luminosity 
was ${\cal L}=126.0 \pm 3.3\pb^{-1}$.


\section{Theoretical predictions}
\label{sec-theo}
Heavy quark production in DIS has been calculated at next-to-leading order
($O(\alpha_s^2)$)
 in the so-called fixed flavour number
 scheme (FFNS) in which only light flavours are present in the proton and
 heavy quarks are produced in the interaction~\cite{np:b374:36}.
The results of this analysis
have been compared to NLO calculations performed with
the {\sc Hvqdis} program~\cite{np:b452:109,pl:b353:535}.
The  renormalisation and factorisation scales 
were set to $\mu_R^2=\mu_F^2=Q^2+4m_q^2$ and the quark masses to $m_c=1.5\gev$ and $m_b=4.75\gev$. The PDFs were obtained by repeating the  ZEUS-S~\cite{pr:d67:012007} PDF fit in the FFNS with quark masses set to the same values as in the
{\sc Hvqdis} calculation.

To calculate muon observables,
the partonic results were interfaced to a model
of HQ fragmentation into weakly decaying heavy hadrons and
of the decay of heavy hadrons into muons.
The hadron momentum was obtained by scaling the 
quark momentum according to the fragmentation function of Peterson
et al.~\cite{pr:d27:105} with
the parameter  $\epsilon_c=0.055$ for charm and $\epsilon_b=0.0035$ for beauty.
This choice of $\epsilon_c$ corresponds to
$\epsilon_c=0.035$ for $D^*$ mesons~\cite{np:b565:245}
since kinematic considerations~\cite{jhep:0604:006}
and direct measurements~\cite{pr:d73:032002}
show that, on average, the momentum of  
the weakly decaying hadrons is $\approx 5\%$ lower than that of $D^*$ mesons.

The semileptonic decay spectrum for charm was taken from a recent CLEO
measurement~\cite{prl:97:251801}.
The decay spectrum for beauty hadrons was taken from
the {\sc Pythia}~\cite{cpc:135:238} Monte Carlo (MC),
mixing direct SL decays and cascade decays through charm
according to the measured branching ratios~\cite{jp:g33:1}.
It was checked that the MC described BELLE and BABAR
data~\cite{pl:b547:181,*pr:d67:031101} well.
The branching ratios were set to ${\cal B}(c\rightarrow \mu)=0.096\pm0.004$ and
${\cal B}(b\rightarrow \mu)=0.209\pm0.004$~\cite{jp:g33:1}.

The uncertainty on the theoretical predictions was evaluated by
independently varying $\mu_R$ and $\mu_F$ by a factor two;
by varying the HQ masses simultaneously to $(m_c,m_b)=(1.3,4.5),(1.7,5.0)\gev$ in the calculation and in the PDF fit;
by varying the proton PDFs by their experimental uncertainty and
by varying the fragmentation parameters within
$0.04\!<\!\epsilon_c\!<\!0.12$  (corresponding to  $0.025\!<\!\epsilon_c\!<\!0.085$ for $D^*$ mesons~\cite{newfrag,*h1frag}) and
$0.0015\!<\!\epsilon_b\! <\!0.0055$. 
As a further check, the fragmentation was performed
by scaling the sum of the energy and the momentum parallel to the HQ direction, $E\!+\!p_{||}$, rather than the HQ momentum. The total theoretical uncertainty was obtained by adding in quadrature the effects of each variation. In the beauty case, the total uncertainty is dominated by the variation of $\mu_R$ and of the mass while for charm the variation of $\epsilon_c$ also gives a large contribution.  

The calculations of $F_2^{c\bar{c}}$ and $F_2^{b\bar{b}}$ in the FFNS
were performed using {\sc Hvqdis} and cross checked with 
the QCD evolution code~\cite{upub:botje:qcdnum1612} used in the ZEUS PDF fit.


\section{Monte Carlo samples}
\label{sec-mc}

Charm and beauty MC samples were generated using
 {\sc Rapgap 3.00}~\cite{cpc:86:147}
 to simulate the leading order BGF process.
Parton shower techniques were used to simulate higher order QCD effects. 
Higher order QED effects were included through  {\sc Heracles 4.6}~\cite{cpc:69:155}.
The CTEQ5L~\cite{epj:c12:375} PDFs were used and the HQ
masses were set to $m_c=1.5\gev$ and $m_b=4.75\gev$.

Light flavour MC events were extracted from an inclusive DIS sample
generated with {\sc DjangoH 1.3}~\cite{proc:hera:1991:1419,*spi:www:djangoh11}
which  is interfaced to {\sc Lepto 6.5}~\cite{cpc:101:108}
to simulate the hadronic final state with the matrix element plus parton shower ({\sc meps}) model and to {\sc Heracles 4.6} to include 
 electroweak radiative corrections.  
The CTEQ5D~\cite{epj:c12:375} parton density was used.

Inelastic $J/\psi$ production was simulated with
{\sc Cascade}~\cite{epj:c19:351}
since that model generally describes the DIS data of a previous
publication~\cite{epj:c44:13}.

The above samples corresponded to at least five times the luminosity
of the data. A smaller light quark sample was generated with {\sc Rapgap}
and mixed with the heavy quark {\sc Rapgap} samples for the study of the
inclusive DIS control sample (Section~\ref{sec-fit}).

Fragmentation and particle decays were simulated
using the {\sc Jetset/Pythia} model~\cite{cpc:82:74,cpc:135:238}.
The lepton
energy spectrum from charm decays was reweighted to agree with
CLEO data~\cite{prl:97:251801}. 
The MC events were passed through a full simulation of the 
ZEUS detector based on {\sc Geant 3.21}~\cite{tech:cern-dd-ee-84-1}.
They were then subjected to the same trigger criteria and reconstructed 
with the same programs as used for the data.

\section{Experimental set-up}
\label{sec-exp}

\Zdetdesc

Charged particles were tracked in the silicon microvertex detector (MVD)~\citeMVD and in the
central tracking detector (CTD)~\citeCTD,
which operated in a magnetic field of $1.43\Tesla$ provided by a thin 
superconducting solenoid.
 The MVD consisted of a barrel (BMVD)
 and a forward (FMVD) section  with three cylindrical layers
 and four vertical planes of single-sided silicon detectors, respectively.
 The CTD consisted of 72~cylindrical drift chamber 
layers, organised in 9~superlayers covering the polar-angle region 
\mbox{$15^\circ<\theta<164^\circ$}. After alignment, the single-hit resolution of the BMVD was $25$~$\mu$m and the impact parameter resolution of the CTD--BMVD system for high-momentum tracks was $\approx 100$~$\mu$m.

The high-resolution uranium--scintillator calorimeter (CAL)~\citeCAL consisted of three parts: the forward (FCAL), the barrel (BCAL) and the rear (RCAL) calorimeters. Each part was subdivided transversely
into towers
and longitudinally into one electromagnetic section and either one (in RCAL)
or two (in BCAL and FCAL)   hadronic sections. Under test-beam conditions, the CAL single-particle relative
energy resolutions were $\sigma(E)/E=0.18/\sqrt{E}$ for leptons and $\sigma(E)/E=0.35/\sqrt{E}$ for
hadrons, with $E$ in $\Gev$.
The energy of electrons hitting the RCAL was corrected for the presence of dead material using the rear presampler detector (PRES)~\cite{nim:a382:419} and the
small angle rear tracking detector (SRTD)~\cite{nim:a401:63}.

The muon system consisted of rear/barrel (R/BMUON) \cite{nim:a333:342} and forward (FMUON) \cite{zeus:1993:bluebook} tracking detectors.
The B/RMUON consisted of limited-streamer (LS) tube chambers
placed behind the BCAL (RCAL), inside and outside a magnetised iron yoke
surrounding the CAL.
The barrel and rear muon chambers cover polar angles from $34^{\rm o}$
to $135^{\rm o}$ and from $135^{\rm o}$ to $171^{\rm o}$, respectively.
The FMUON consisted of six trigger planes of LS tubes
and four planes of drift chambers
covering the angular region from $5^{\rm o}$ to $32^{\rm o}$.
The muon system exploited the magnetic field of the iron yoke and,
in the forward direction, of two iron toroids magnetised to $\approx 1.6$~T
to provide an independent measurement of the muon momentum.

The luminosity was measured using the Bethe-Heitler reaction $ep \rightarrow e\gamma p$
with the luminosity detector which consisted of two independent systems, a
photon calorimeter~\cite{Desy-92-066,*zfp:c63:391,*acpp:b32:2025} and a magnetic spectrometer~\cite{physics-0512153}.


\section{Event reconstruction and selection}
\label{sec-selection}

A three-level trigger was used to select events online~\cite{zeus:1993:bluebook,nim:a580:1257,*uproc:chep:1992:222}.
DIS events were selected 
by requiring a scattered electron in the CAL.

A scattered electron with energy $E'_e>8\gev$ was required offline.
The primary vertex had to be within $\pm 30$~cm in $Z$ 
from the nominal interaction point. 

Muons were reconstructed by matching a CTD+MVD track 
to a track segment in the inner or outer B/RMUON chambers or to an FMUON
track crossing at least four FMUON planes. This B/RMUON selection was 
looser than in some previous analyses, which required
the muons to reach the external
chambers~\cite{pl:b599:173,pr:d70:012008},
allowing a lower  threshold for the muon transverse momentum.

The central track associated to a B/RMUON candidate
was required to pass at least three CTD
superlayers and to have at least four hits in the MVD to allow a good impact
parameter measurement. The tracks associated to FMUON candidates were
required to pass at least one CTD superlayer, corresponding to
at least four degrees of freedom in the track fit.

Muons were accepted in the kinematic region defined by
$$p_T^{\mu}>1.5\gev, -1.6<\eta^{\mu}<2.3.$$ 

The hadronic system (including the muon) was reconstructed
from energy flow objects (EFOs)~\cite{thesis:briskin:1998} that combine the information from calorimetry and tracking, corrected for energy loss in the dead
material. The EFOs were corrected using the measured momenta
of identified muons~\cite{pr:d70:012008,thesis:turcato:2002}.
A reconstructed four-momentum ($p^i_X,p^i_Y,p^i_Z,E^i$) was assigned to each EFO $i$.

To select a clean DIS sample, the following cuts on global variables were applied: \begin{eqnarray*}
   (E\!-\!P_Z)_{\rm tot} =&  (E\!-\!P_Z)_h +E'_e (1\!-\!\cos\theta_e)& \subset [40,80] \gev  \\
   y_e         =& 1- E'_e (1\!-\!\cos\theta_e)/(2E_e) &<0.7 \\
   y_{\rm JB}   =& (E\!-\!P_Z)_h/(2E_e) &> 0.01    \\
   Q^2_{\Sigma} =& (E'_e \sin \theta_e)^2/(1-y_{\Sigma})& > 20 \gev^2, 
\end{eqnarray*}
where $(E\!-\!P_Z)_h=\sum_{i \subset \rm EFOs} E^i\!-\!p^i_Z$, 
$y_{\Sigma}=(E\!-\!P_Z)_h/(E\!-\!P_Z)_{\rm tot}$~\cite{nim:a361:197},
and $\theta_e$ is the electron  polar angle. 
These cuts restricted the accessible inelasticity $y=Q^2/(xs)$ and $Q^2$ to $0.01<y<0.7$ and $Q^2>20\gev^2$.
The DIS variables $x$ and $Q^2$ were reconstructed using the $\Sigma$
estimators $Q^2_{\Sigma}$ and $x_{\Sigma}=Q^2_{\Sigma}/(s  y_{\Sigma})$~\cite{nim:a361:197}.

To remove background events with isolated muons
($\gamma\gamma \rightarrow \mu^+ \mu^-$, $J/\psi$ and $\Upsilon$ decays)
and  residual cosmic muons,
an anti-isolation cut was applied
by requiring that the hadronic energy in a cone of radius $1$ in the $\eta-\phi$
plane around the muon candidate, excluding the muon itself, was $E^{\rm iso}>0.5\gev$. From MC studies this cut was 98\% (90\%) efficient for charm
(beauty).

Jets were reconstructed from EFOs
using the $k_T$ algorithm \cite{np:b485:291} in the longitudinally invariant
 mode~\cite{np:b406:187,*pr:d48:3160}.
About 96\% of the muon candidates were associated to
a jet with transverse momentum
(including the muon)  $p_T^{\rm jet}>2.5\gev$ 
 and kept for further analysis.

After the above selection, the final sample contained 11126 muons.
A subsample of 35 events with more than one muon was found, 28 of which
consisted of $\mu^+\mu^-$ pairs.
A $J/\psi$ signal of 9 events was observed in the ${\mu^+\mu^-}$
invariant mass distribution.
The total contamination from $J/\psi$ production was estimated with the
{\sc Cascade} MC, normalised to the observed
$J/\psi$ signal. It was found to be $(0.9\pm0.3)\%$ and was neglected
 in the analysis. 


\section{Extraction of the charm and beauty fractions}
\label{sec-fit}
The sample of selected muon candidates contained
signal muons from charm and beauty decays
and background from in-flight
 $\pi^{\pm}$ and $K^{\pm}$ decays
and from the punch through of hadronic jets in the muon chambers. 
Candidates from  in-flight  decays
and punch through,
which are subsequently denoted
as ``false muons", were present both in the 
LF events and in events containing HQs.

The fractions of muons originating from charm, beauty or LF
events were determined from a simultaneous fit of 
three discriminating variables sensitive to different aspects of HQ decays:
\begin{itemize}
\item $p_T^{\rm rel}$, the muon momentum component
transverse to the axis of the associated jet, 
$p_T^{\rm rel}=|\mathbf{p}^{\:\mu} \times \mathbf{p}^{\rm \:jet }|/|\mathbf{p}^{\rm \: jet}|$. Due to the large $b$ mass, muons from beauty hadron
decays have a harder  $p_T^{\rm rel}$ spectrum than those from
charm or light quarks; 
\item $\delta$, the distance of closest approach of the muon track
 to the centre of the interaction region (beam spot) in the $X,Y$ 
plane.
A positive sign was assigned to $\delta$ if the muon track crosses the axis of
the associated jet in the jet hemisphere, negative otherwise.
The beam spot position was obtained by fitting the reconstructed primary vertex
distribution for every 2000 $ep$ events. The size of the interaction region was
$80\! \times\! 20$~$\mu$m$^2$ in $X\! \times\! Y$.
Muons from decays of long-lived heavy quarks tend
to have positive $\delta$ while tracks originating from the
primary interaction have a  symmetric  $\delta$ distribution around zero,
 corresponding to the experimental resolution. 
\item $p_T^{\rm miss || \mu}$, the missing transverse momentum parallel to the muon direction. The missing transverse momentum vector was calculated
using the electron and the EFOs.
The $p_T^{\rm miss || \mu}$ distribution has a positive tail of
events containing  semileptonic HQ decays due to the
presence of the neutrino.
\end{itemize}

A control sample of inclusive DIS data, selected similarly
to the muon sample but without any muon requirement,
was used to test the quality of the simulation of these variables.
The control sample is dominated by LF events, containing, according to MC,
about 18\% (1\%) of $c$ ($b$) events. The $p_T^{\rm rel}$
distribution of inclusive tracks in the
control sample was reasonably well reproduced by both the {\sc DjangoH} and
the {\sc Rapgap} inclusive DIS samples. The small differences
(at most 10\% at $p_T^{\rm rel}>2\gev$) were corrected for by applying a
bin-by-bin  correction to the $p_T^{\rm rel}$ distribution of the LF and charm
MC samples similarly to a previous publication~\cite{pr:d70:012008}.
The quality of the MC description of
$p_T^{\rm miss || \mu}$ was also evaluated in the control
sample by studying a similar $p_T$-balance variable:
the missing transverse momentum parallel to the electron $p_T^{\rm miss || e}$.
The best description of the  $p_T^{\rm miss || e}$ distribution of the
inclusive DIS sample  was obtained by
shifting the hadronic transverse momentum by $(0.1\pm0.1)$~GeV in the MC
and by increasing the hadronic transverse momentum resolution by $(5\pm5)\%$
in the case of {\sc Rapgap} and by $(0\pm5)\%$ in the case of {\sc DjangoH}. 
The resolution on $\delta$ was studied using tracks in the inclusive
DIS sample. Since it was underestimated in the MC by  
$\approx 15\%$, a $p_T$-dependent smearing~\cite{thesis:miglioranzi:2006}
was applied to the MC, similarly to what was done in a
previous publication~\cite{newbphp}.

The fractions of $b$, $c$ and LF events were obtained by fitting
a combination of MC distributions to the measured
three-dimensional distribution of the discriminating variables
~\cite{thesis:bindi:2008}.
The fit range was $|\delta|<0.1$~cm, $p_T^{\rm rel}<2.5\gev$ and $|p_T^{\rm miss || \mu}|<10 \gev$.
A precise measurement of $\delta$ was only possible inside
the region covered by the BMVD. Hence for events 
with muons reconstructed in the FMUON (4\% of the total) only
 $p_T^{\rm rel}$ and $p_T^{\rm miss || \mu}$ were used in the fit.
A Poisson likelihood fit was used, taking into account
the limited MC statistics.

\begin{figure}[p]
\begin{center}
\psfig{figure=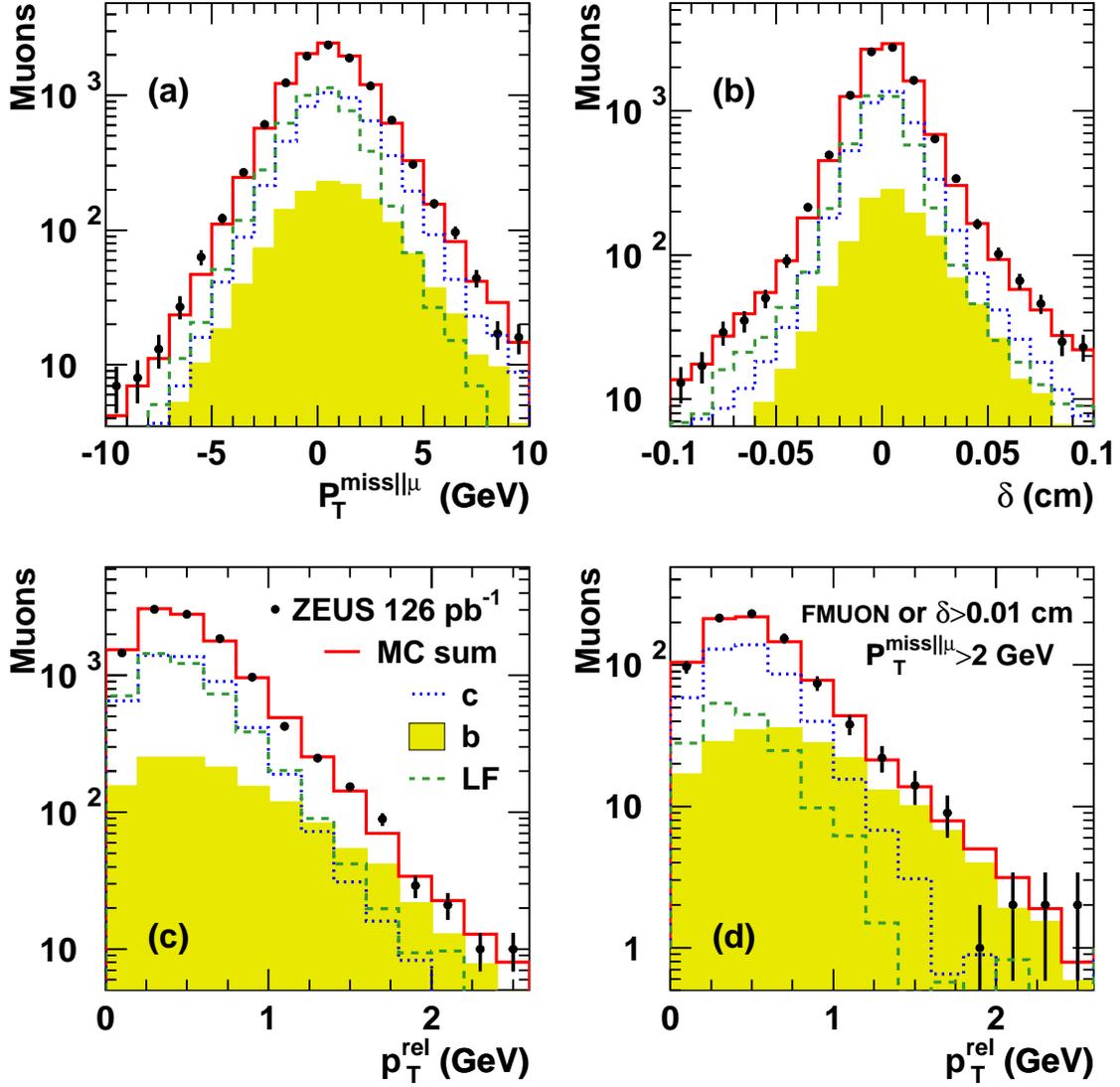,width=17. cm,clip=}
\end{center}
\caption{Distributions of (a) $p_T^{{\rm miss}||\mu}$, (b) $\delta$, (c) $p_T^{\rm rel}$ for the selected sample of muons in DIS, and of (d) 
$p_T^{\rm rel}$ for a signal-enriched subsample with $p_T^{{\rm miss}||\mu}>2$~GeV and either a muon in FMUON or $\delta>0.01$~cm. The data (points)
are compared to the MC expectation (solid line) with the normalisation of the $c$ (dotted line), $b$ (shaded histogram)  and light flavours (dashed line), LF, components obtained from the global fit. The error bars
correspond to the square root of the number of entries.}
\label{f2}
\vfill
\end{figure}

\begin{figure}[p]
\vfill
\begin{center}
\psfig{figure=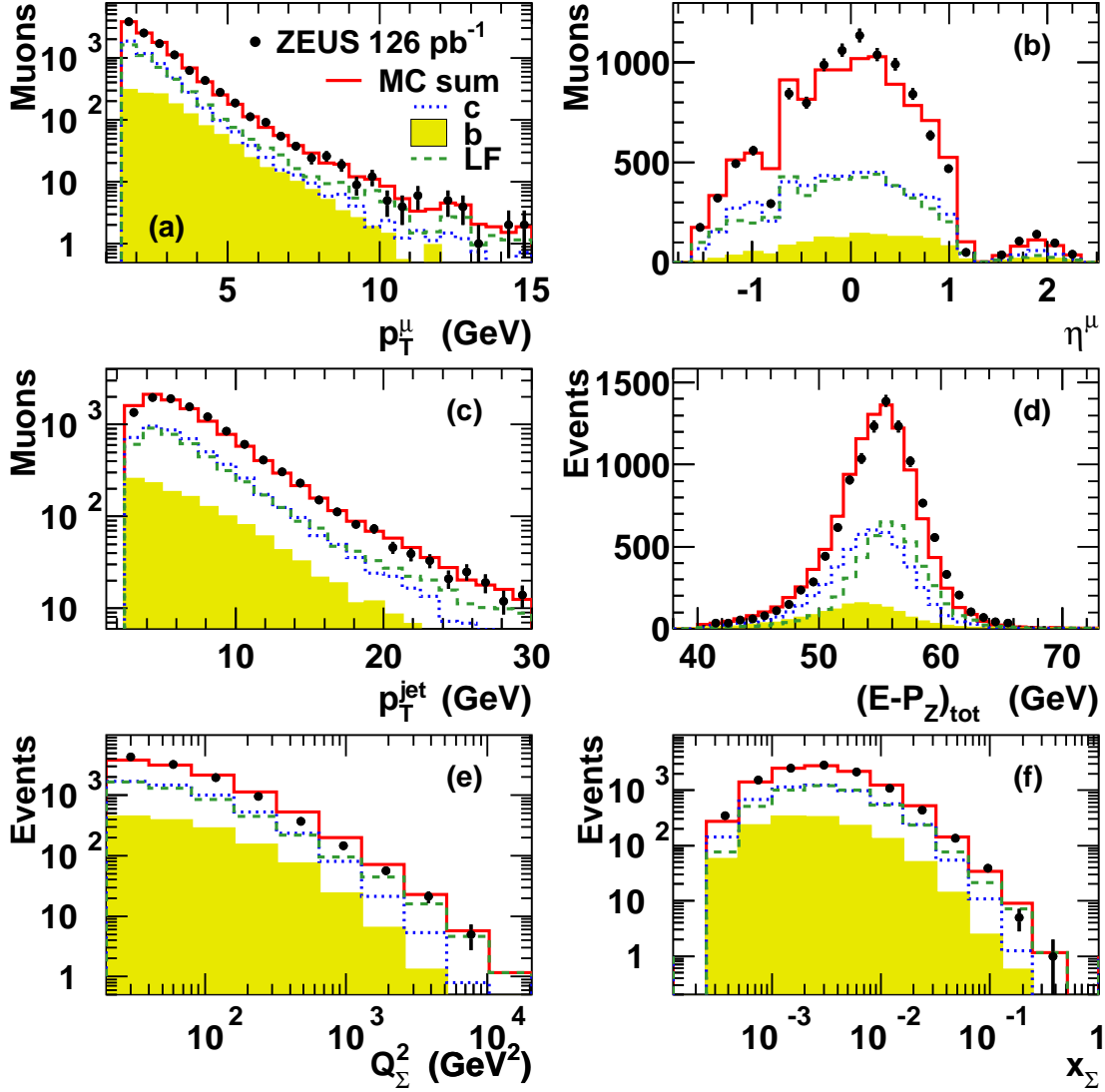,width=17. cm,clip=}
\end{center}
\caption{Distributions of
 (a) $p_T^{\mu}$, (b) $\eta^{\mu}$, (c) $p_T^{\rm jet}$, (d) $(E\!-\!P_Z)_{\rm tot}$, (e) $Q^2_{\Sigma}$ and (f)  $x_{\Sigma}$
 for the selected sample of muons in DIS.
 The data (points) are compared to the MC expectation with the normalisation
 of the $c$, $b$ 
and light flavours, LF, components obtained from the global fit.}
\label{f1}
\vfill
\end{figure}

The global charm and beauty fractions resulting from the fit were
$$ f_c=0.456 \pm 0.029 ({\rm stat.}) ;\,\,\,  f_b= 0.122 \pm 0.013 ({\rm stat.}) $$
with a correlation coefficient $\rho_{cb}=-0.43$. 
Figure \ref{f2}(a-c) shows the distributions of the three discriminating
variables compared to the MC distributions with the normalisation corresponding
to the fit.  While $\delta$ and $p_T^{\rm miss || \mu}$ provide discrimination between LF and HQs, $p_T^{\rm rel}$ discriminates between beauty and the 
other components.
Figure 1(d) shows the distribution of $p_T^{\rm rel}$ for a signal-enriched subsample.
The distributions of $p_T^{\mu}$, $\eta^{\mu}$, $p_T^{\rm jet}$, $E-P_Z$, $Q^2_{\Sigma}$ and $x_{\Sigma}$ for the data and for the MC samples normalised
according to the fit are shown in Fig.~\ref{f1}. The overall agreement is
satisfactory.


\section{Acceptance and QED corrections}
\label{sec-measurement}

The visible cross sections for muons from charm and beauty decays,
including beauty cascade decays via $c,\bar{c},\tau$ and $\psi$,
were measured in the kinematic range
\begin{equation}
Q^2>20\gev^2;\; 0.01<y<0.7;\; p_T^{\mu}>1.5\gev;\; -1.6<\eta^{\mu}<2.3.
\label{e1}
\end{equation}
The cross sections were calculated using
 $$ \sigma^{q} = \frac{f_{q} \, N}{A_{q} \,{\cal L}}\, C_{r},$$
where  $f_{q}$ is the HQ fraction from the fit, $N$ is the number of reconstructed muons, $A_{q}$ is the acceptance, $C_{r}$ is the QED radiative correction, and $q=c,b$. Differential cross sections 
were measured by repeating the fit in bins of the reconstructed variable $V$
as $d\sigma/dV = \sigma^{q}_i/\Delta V_i$, 
where $\sigma^{q}_i$ is the cross section in the bin and $\Delta V_i$ is the bin width.

The acceptance $A_{q}$ was evaluated from the
MC simulation as the number of reconstructed muons
divided by the number of true muons from decays of the quark $q$.
This definition takes into account the charm and beauty events
in which a ``false muon'' is reconstructed rather than
a signal muon from a HQ decay.
The acceptance included the efficiency of muon reconstruction 
(which in turn includes the efficiency of the muon chambers and of the matching
with central tracking) that was evaluated from an independent exclusive
dimuon sample as explained in previous publications~\cite{pr:d70:012008,thesis:turcato:2002}. The muon reconstruction efficiency was around
$50\%$ for central muons with  $p_T^{\mu}>2\gev$.
The acceptance for $c$ ($b$) ranged from $23\%$ ($16\%$) at  $1.5<p_T^{\mu}<2.5\gev$ to $\approx 35\%$ ($25\%$) at $p_T^{\mu}>2.5 \gev$.
The difference in acceptance between $c$ and $b$ was mainly due to
the different contribution from ``false muons''
which was $\approx 25\%$ for $c$ and $\approx 3\%$ for $b$.

According to the MC simulation, the probability to find a ``false muon'' in a DIS event
(before any muon selection) is ${\cal P}_{\rm false}\approx 0.1\%$,
almost independently from the event being $c$, $b$ or LF.
The  ability of the MC to reproduce ${\cal P}_{\rm false}$ was studied
by comparing the number of LF events in the data, as obtained from the fit,
to the absolute prediction by {\sc DjangoH}.
The data/MC ratio, ${\cal P}_{\rm false}^{\rm data}/{\cal P}_{\rm false}^{\rm MC}$,
was estimated as $0.80 \pm 0.20$ in the RMUON, $1.10 \pm 0.20$ in  the BMUON, and $1.05 \pm 0.40$ in the FMUON,
in agreement with previous studies~\cite{thesis:longhin:2003}.

The cross sections were corrected to the QED Born level,
calculated using a running coupling costant $\alpha_{\rm em}$,
such that they can
be compared directly to the QCD predictions from the {\sc Hvqdis} program (Section~\ref{sec-theo}).
The radiative corrections were obtained as
 $C_{r}=\sigma_{\rm Born}/\sigma_{\rm rad}$,
where $\sigma_{\rm Born}$ is the
 {\sc Rapgap} cross section with the QED corrections turned off but 
keeping $\alpha_{\rm em}$ running and 
 $\sigma_{\rm rad}$ is
the {\sc Rapgap} cross section with the full QED corrections, as in the standard
MC samples.
The corrections  were typically $C_r\approx 1.05$ and at maximum $1.10$ in the highest $Q^2$ bin.


\section{Systematic uncertainties}
\label{sec-syst}
The following systematic uncertainties were considered (the effects on the total
visible cross section for $c$ and for $b$ is given in parentheses):
\begin{enumerate}
\item B/RMUON efficiency: it was varied by its uncertainty of on average $\pm 5\%$ $(\mp 5, \mp 5)\%$;
\item FMUON  efficiency: it was varied by $\pm 20\%$  $(\mp 2, \mp 5)\%$;
\item ``false muon'' probability: it was varied within the corresponding uncertainty for each muon detector $(^{-3}_{+4},\mp 1)\%$;
\item global energy scale: it was varied by
$\pm 2\%$ $(^{-4}_{+5}, ^{-3}_{+2})\%$;
\item calibration of $p_T^{{\rm miss}||\mu}$: it was evaluated by varying the hadronic transverse momentum in the MC by $\pm 0.1$~GeV,
 as allowed by  the transverse momentum balance in the control
 sample $(\pm 12, ^{-2}_{+1})\%$;
\item hadronic energy resolution: it was varied in the MC by $\pm 5\%$
 as allowed by the transverse momentum balance in the control sample  $(^{+1}_{+2}, \mp 7)\%$;
\item simulation of the tails of $p_T^{{\rm miss}||\mu}$: the fits were redone
 in the  restricted range $|p_T^{\rm miss||\mu}|< 5 \gev$  $(0,-6)\%$;
\item resolution on $\delta$:
 the smearing applied to the MC was varied by $\pm 25\%$ as
 allowed by the control sample $(^{-3}_{+2}, ^{+11}_{-9})\%$;
\item $p_T^{\rm rel}$ shape of LF and charm: it was evaluated by varying the
  $p_T^{\rm rel}$ correction by $\pm 50\%$ $(\mp 1.5, ^{+8}_{-5})\%$;
\item hadronic energy flow near the muon: it
 was evaluated by varying the
 cut on $E^{\rm iso}$ by $^{+0.50}_{-0.25}\gev$ $(0, ^{-1}_{~\:0})\%$;
\item jet description: the cut on $p_T^{\rm jet}$ was varied by $\pm 0.5 \gev$ $(\pm 2.5, ^{-3.5}_{+2.5})\%$;
\item charm SL decay spectrum: the reweighting to the CLEO model
was varied by $\pm 50\%$, $(^{-4}_{+3}, ^{+3}_{-2})\%$;
\item MC model dependence: the {\sc Rapgap} $c$ and $b$ samples were
 reweighted to reproduce the corresponding measured differential cross sections
 in $Q^2$ or in $p_T^{\mu}$ and the largest deviation from the nominal
 cross section was taken $(+6,+20)\%$;
\item higher order effects: this uncertainty was evaluated by varying
 the HQ distribution before parton showering in {\sc Rapgap} 
 by the difference between NLO and leading order, as evaluated with
 {\sc Hvqdis} $(^{+6}_{-10},^{+2}_{-3})\%$;
\item MVD efficiency: the efficiency of the cut on the number of MVD
 hits of $(90\pm 3)\%$ was varied by its uncertainty $(\mp 3,\mp 3)\%$;
\item CTD simulation:  tracks were required to pass $\ge 4$ superlayers
 in the B/RMUON region and to have $\ge 7$ degress of freedom in the
 FMUON region $(+1,0)\%$; 
\item integrated luminosity: measurement uncertainty $(\mp 2.6,\mp 2.6)\%$.
\end{enumerate}
The above uncertainties were summed in quadrature to obtain the total systematic uncertainty $(^{+18}_{-19},^{+28}_{-17})\%$.


\section{Cross sections}
\label{sec-results}

The visible cross sections for muons from charm and beauty
decays  in the kinematic region  of Eq.~(\ref{e1}) are
\begin{eqnarray*}
\sigma^{c}=  164  \pm 10 {\rm(stat.)}\: ^{+30}_{-31} {\rm(syst.)~ pb}\\
\sigma^{b}= \:\: 63   \pm \:\: 7 {\rm(stat.)}\: ^{+18}_{-11} {\rm(syst.)~ pb}
\end{eqnarray*} 
to be compared with the NLO QCD cross sections obtained with {\sc Hvqdis}
of $\sigma^{c}=  184  ^{+26}_{-40}$~pb and  $\sigma^{b}=  33 \pm 5$~pb.
The agreement is good for charm while the beauty
cross section is  $2.3$ ($1.9$) standard deviations above the central (upper)
 {\sc Hvqdis} result.
The visible cross sections are a factor  $1.04$ and $2.27$ higher
than the {\rm Rapgap} MC predictions for $c$ and $b$, respectively.

\begin{figure}[p]
\vfill
\begin{center}
\psfig{figure=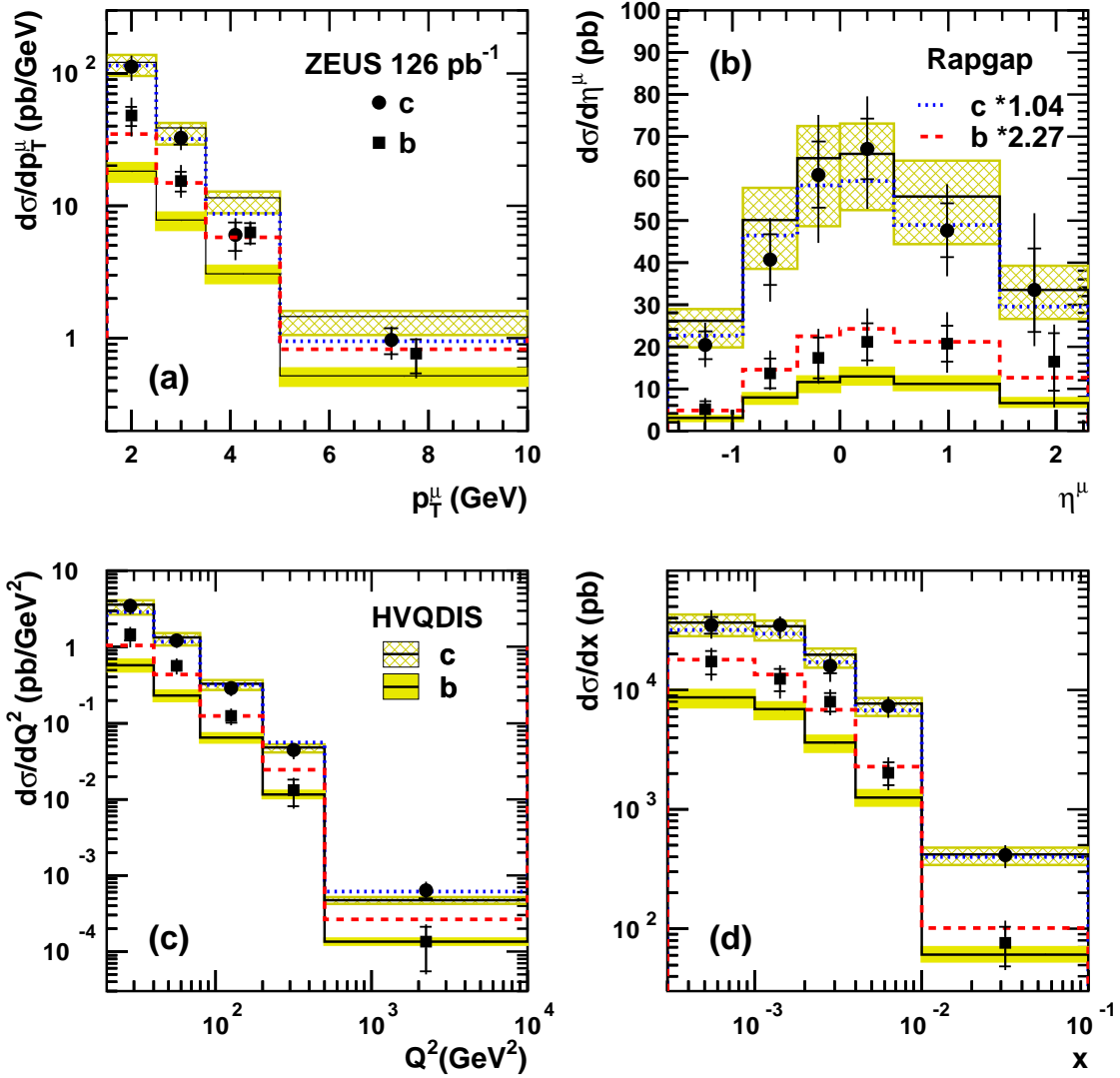,width=16.8 cm,clip=}

\end{center}
\caption{Differential muon cross sections for $c$ and $b$ as a function of (a) $p_T^{\mu}$, (b) $\eta^{\mu}$, (c) $Q^2$, and (d) $x$. The inner error bars show the statistical uncertainty while the outer error bars show the systematic and statistical uncertainties added in quadrature. The bands show the NLO QCD predictions obtained with 
 the {\sc Hvqdis} program and the corresponding uncertainties. 
 The differential cross sections from {\sc Rapgap}, scaled by the factors corresponding to the
 result of the global fit ($1.04$ for $c$ and $2.27$ for $b$), are also shown.
}
\label{f3}
\vfill
\end{figure}

The differential cross sections
as a function of $p_T^{\mu}$, $\eta^{\mu}$, $Q^2$, and $x$
are presented in Table~\ref{t1} and 
compared  in Fig.~\ref{f3} to the NLO QCD predictions based on {\sc Hvqdis}.
The {\sc Rapgap} MC predictions are
also shown, normalised according to the result of the global fit.
Charm and beauty cross sections are similar for $p_T^{\mu}>3.5\gev$.

The charm cross sections are in good agreement with the {\sc Hvqdis}
calculations. 
The tendency of the beauty cross section
to lie above the central NLO prediction is
concentrated at low $p_T^{\mu}$ and low $Q^2$.
The  statistical significance of the difference between the data and the NLO
predicitons is similar to that obtained for the total visible cross
section since the uncertainties are dominated by correlated systematics. 

Both NLO calculations and the {\sc Rapgap} MC give in general a good
description of the shape of the differential cross sections.
The $Q^2$ distributions for beauty is somewhat steeper than predicited by {\sc Rapgap} and {\sc Hvqdis}.
%

\section{Extraction of $F_2^{q\bar{q}}$}
\label{sec-f2}

The heavy quark contribution to the proton structure functions, $F_2^{q\bar{q}}$,
$F_L^{q\bar{q}}$ and the reduced cross section $\tilde \sigma^{q\bar{q}}$ 
are defined in analogy with the inclusive case from the
double differential cross section in $x$ and $Q^2$
for the production of the quark $q$:
$$\frac{d^2\sigma^{q\bar{q}}}{dx\, dQ^2}
 = {\cal K}~ \left[ F_2^{q\bar{q}} (x,Q^2) - \frac{y^2}{Y_+} F_L^{q\bar{q}} 
(x,Q^2) \right]
 = {\cal K}~ \tilde \sigma^{q\bar{q}}(x,Q^2,s),$$
where $ {\cal K}=Y_+ (2\pi \alpha_{\rm em}^2)/(x Q^4)$ and $Y_+=1+(1-y)^2$.

The muon cross sections, $\sigma^q$,
measured in bins of $x$ and $Q^2$,  were
used to extract $F_2^{q\bar{q}}$ at a reference points in the $x,Q^2$ plane by:
$$ F_2^{q\bar{q}}(x,Q^2)=\sigma^{q} \frac{F_2^{q\bar{q}, \rm th}(x,Q^2)}{\sigma^{q,{\rm th}}},$$
where $F_2^{q\bar{q}, \rm th}(x,Q^2)$ and $\sigma^{q,{\rm th}}$
were calculated at NLO in the FFNS using the {\sc Hvqdis} program.
The reference points were chosen close to
the average $x$ and $Q^2$ of the events within each bin.
Charm produced in $b$ decays was not included in the definition of $\sigma^{c}$.

This  procedure contains several corrections:
the extrapolation from the restricted muon kinematic range
($p_T^{\mu}>1.5 \gev, -1.6<\eta^{\mu}<2.3$) to the full muon phase space;
the $q \rightarrow \mu$ branching ratio;
the correction for the longitudinal structure function $F_L^{q\bar{q}}$ and
the correction from a bin-averaged cross section to a point
value (bin centring).

The largest uncertainty is related to the
extrapolation to the full muon phase space. The kinematic acceptance, ${\cal A}$,
defined as the fraction of muons from HQ decays that was generated in the
restricted kinematic region is, on average,
 $\langle {\cal A} \rangle=13\%$$(27\%)$, for charm (beauty).
 According to {\sc Hvqdis}, in the charm case, ${\cal A}$ becomes sizeable
 (${\cal A}>0.25 \langle {\cal A} \rangle$)
when one of the two charm quarks in the event has
$p_T> 3\gev$ and its rapidity is in the range $(-1.5 : 2.5)$, which corresponds
to the phase space containing 88\%  of the cross section.
In the beauty case, ${\cal A}$ is sizeable over the full HQ phase space.

The theoretical uncertainty in the extraction of $F_2^{q\bar{q}}$ was evaluated
by varying the  {\sc Hvqdis} parameters as explained in Section~\ref{sec-theo} and by using a different PDF set (CTEQ5F). It is dominated by the fragmentation uncertainty.
As a further check,  $F_2^{q\bar{q}}$ was also evaluated taking ${\cal A}$ from
{\sc Rapgap} and found to be consistent within the quoted uncertainties.
 
\begin{figure}[p]
\vfill
\begin{center}
\psfig{figure=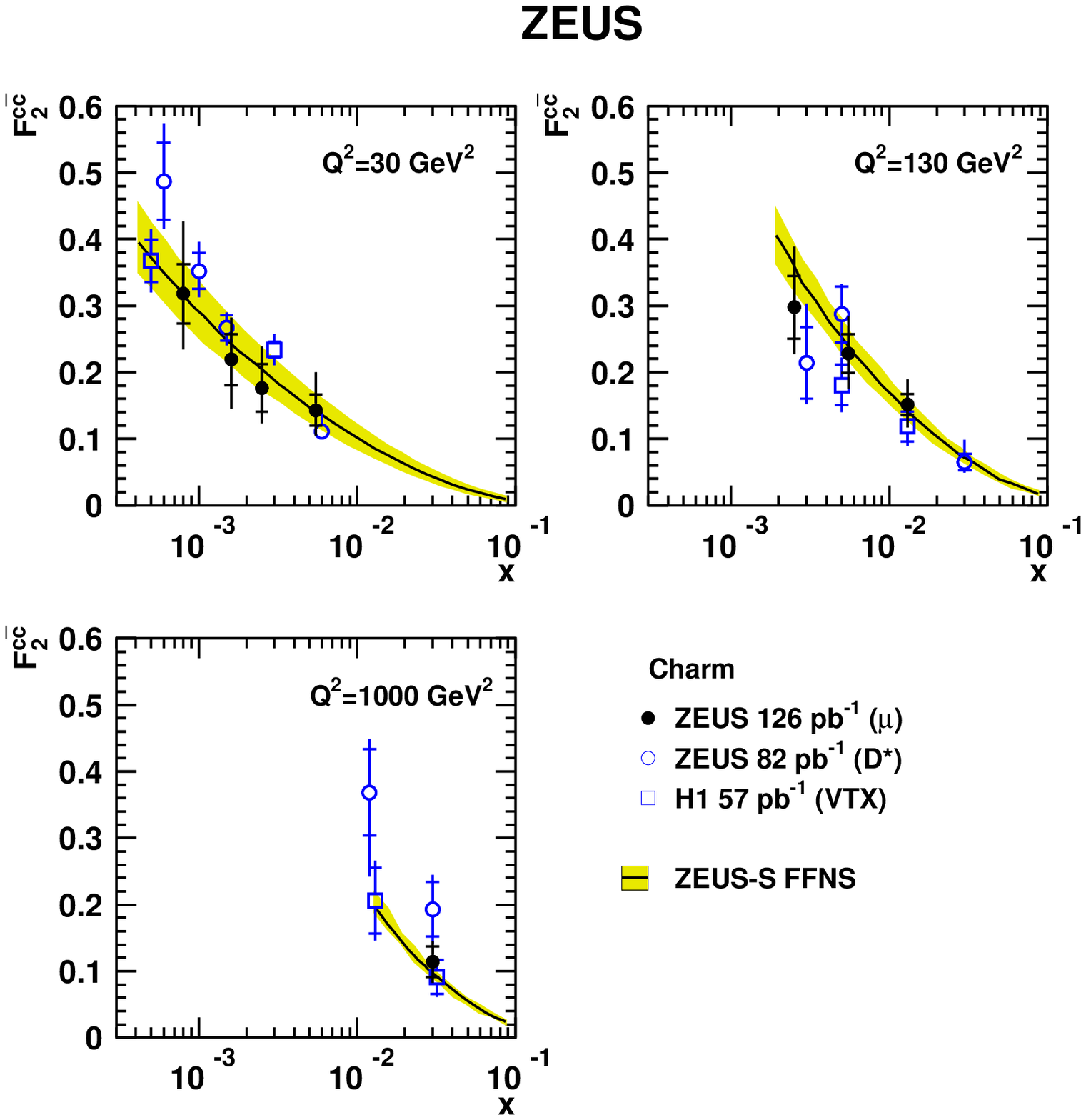,width=17. cm,clip=}

\end{center}
\caption{The structure function $F_2^{c\bar{c}}$ (filled symbols) compared
to previous results (open symbols) and to the NLO QCD  predictions 
in the FFNS using the ZEUS-S PDF fit. 
The inner error bars are the statistical uncertainty while
the outer bars represent the statistical, systematic and theoretical uncertainties added in quadrature. 
The band represents the uncertainty on the NLO QCD
prediction. 
Previous data have been corrected to the reference $Q^2$ values used in this analysis: ZEUS $D^*$ $500\rightarrow 1000\gev^2$;
H1 VTX $25\rightarrow 30$, $200\rightarrow 130$, $650\rightarrow 1000\gev^2$.
}
\label{f4}
\vfill
\end{figure}

\begin{figure}[p]
\vfill
\begin{center}
\psfig{figure=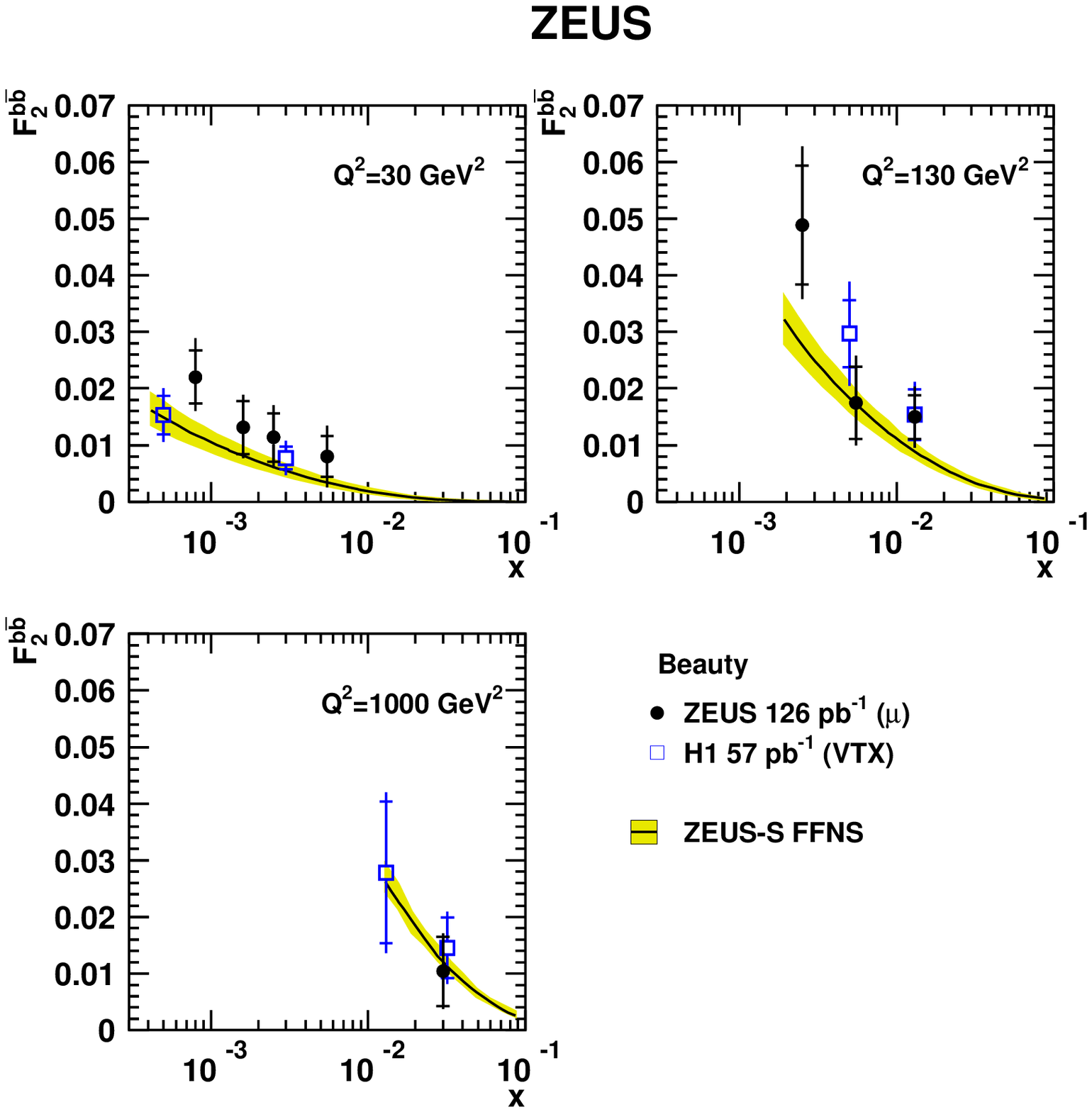,width=17. cm,clip=}
\end{center}
\caption{The structure function $F_2^{b\bar{b}}$ (filled symbols) compared
to previous results (open symbols) and to the NLO QCD  predictions 
in the FFNS using the ZEUS-S PDF fit.
The inner error bars are the statistical uncertainty while
the outer bars represent the statistical, systematic and theoretical uncertainties added in quadrature. 
The band represents the uncertainty on the NLO QCD
prediction. Previous data have been corrected to the reference $Q^2$ values used in this analysis: $25\rightarrow 30$, $200\rightarrow 130$, $650\rightarrow 1000\gev^2$.
}
\label{f5}
\vfill
\end{figure}

The muon cross sections  in bins of $x$ and $Q^2$
are given in Table~\ref{t2}. The extracted $F_2^{c\bar{c}}$ and 
$F_2^{b\bar{b}}$ are presented in Tables~\ref{t3a} and \ref{t3b}
and shown in Figs.~\ref{f4} and ~\ref{f5}. Also given in Tables~\ref{t3a} and \ref{t3b} are the factor ${\cal A}$ and the correction for the longitudinal
structure function $C_L=  \tilde \sigma^{q\bar{q}} / F_2^{q\bar{q}}$
as obtained from the NLO theory.
The effects of the individual sources of systematic and theoretical
uncertainty are given in Tables~\ref{t4} and \ref{t5} in the Appendix
 \footnote{
 The effects of the individual sources of systematic and theoretical
 uncertainty are also available from
{\tt http://www-zeus.desy.de/public\_results/functiondb.php?id=ZEUS-pub-09-003}.
}.

Figure~\ref{f4} also contains a comparison of $F_2^{c\bar{c}}$
with previous results based on the measurement of $D^*$
mesons from ZEUS~\cite{pr:d69:012004} 
and to results from the  H1 collaboration
based on inclusive lifetime tagging (VTX)~\cite{epj:c40:349,*epj:c45:23}.
The previous results were corrected to the $Q^2$ values
of the present analysis, using the NLO theory.
The agreement of the different data sets, obtained with
different charm tagging techniques, is good.
At high $Q^2$, the precision of present data is similar or better than for 
the previous results. 
The NLO QCD calculations  are also shown.

Figure~\ref{f5} shows the extracted $F_2^{b\bar{b}}$ from this analysis and
also a previous H1 result~\cite{epj:c40:349,*epj:c45:23}, corrected to the reference $Q^2$ values used in the present analysis. The two data sets are in good agreement. The precision of the present measurement is similar to that of the H1 data at high $Q^2$. The QCD calculations are also shown.

\begin{figure}[p]
\vfill
\begin{center}
\psfig{figure=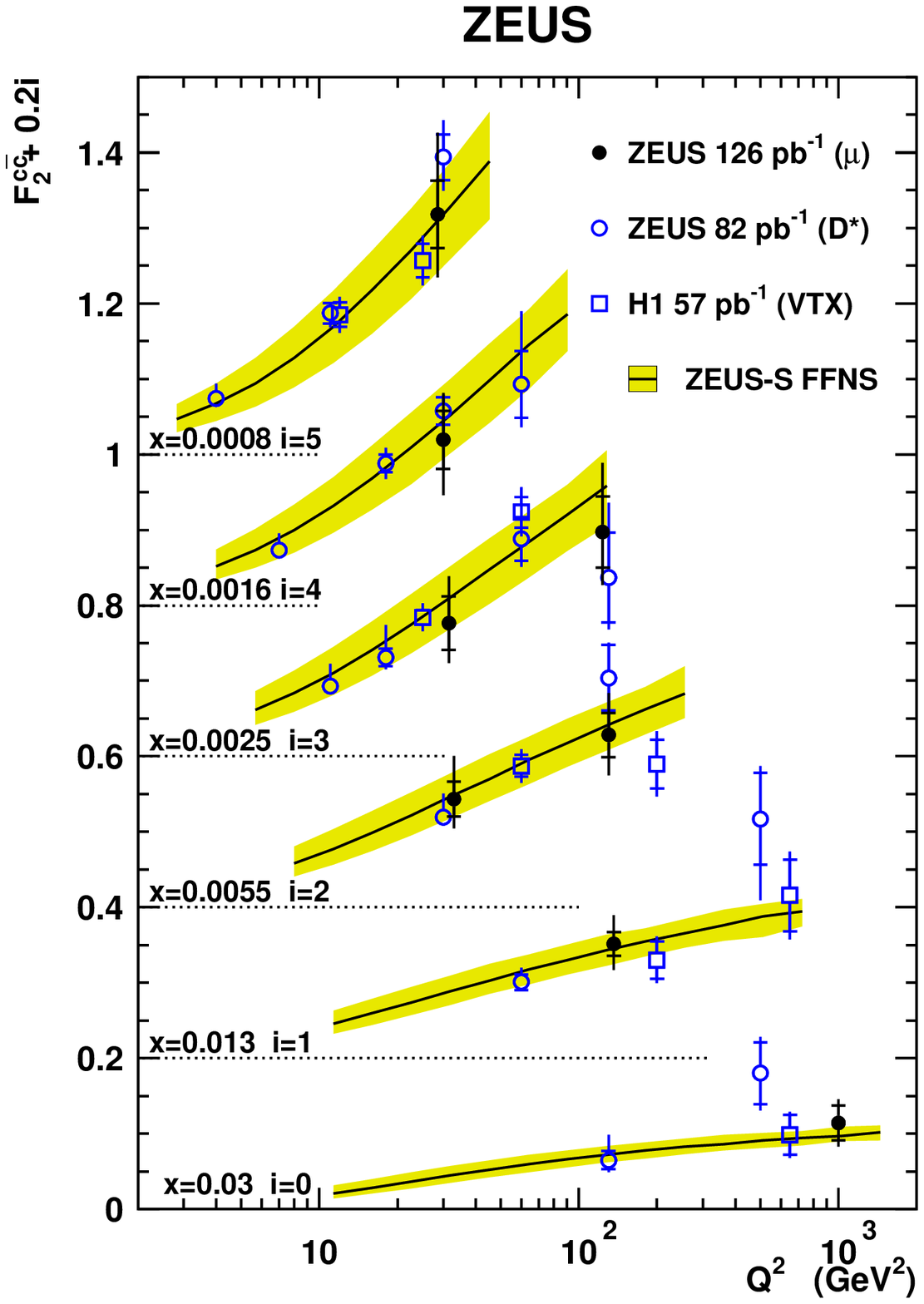,width=13.5 cm,clip=}
\end{center}
\caption{The structure function $F_2^{c\bar{c}}$ (filled symbols) plotted as a function of $Q^2$ for fixed $x$ values.
The curves represent the  NLO QCD predictions 
in the FFNS using the ZEUS-S PDF fit.
The inner error bars are the statistical uncertainty while
the outer bars represent the statistical, systematic and theoretical uncertainties added in quadrature.  The band represents the uncertainty on the NLO QCD
prediction. A selection of previous data (open symbols) is
also shown, corrected to the reference $x$ values used in this analysis:
ZEUS $D^*$:
$0.001  \rightarrow 0.0008$, 
$0.0015 \rightarrow 0.0016$,
$0.003  \rightarrow 0.0025$,
$0.006  \rightarrow 0.0055$,
$0.012  \rightarrow 0.013$;
H1 VTX:
$0.0005 \rightarrow 0.0008$, 
$0.002 \rightarrow 0.0025$,
$0.005 \rightarrow 0.0055$,
$0.032 \rightarrow 0.030$.
}
\label{f6}
\vfill
\end{figure}

\begin{figure}[p]
\vfill
\begin{center}
\psfig{figure=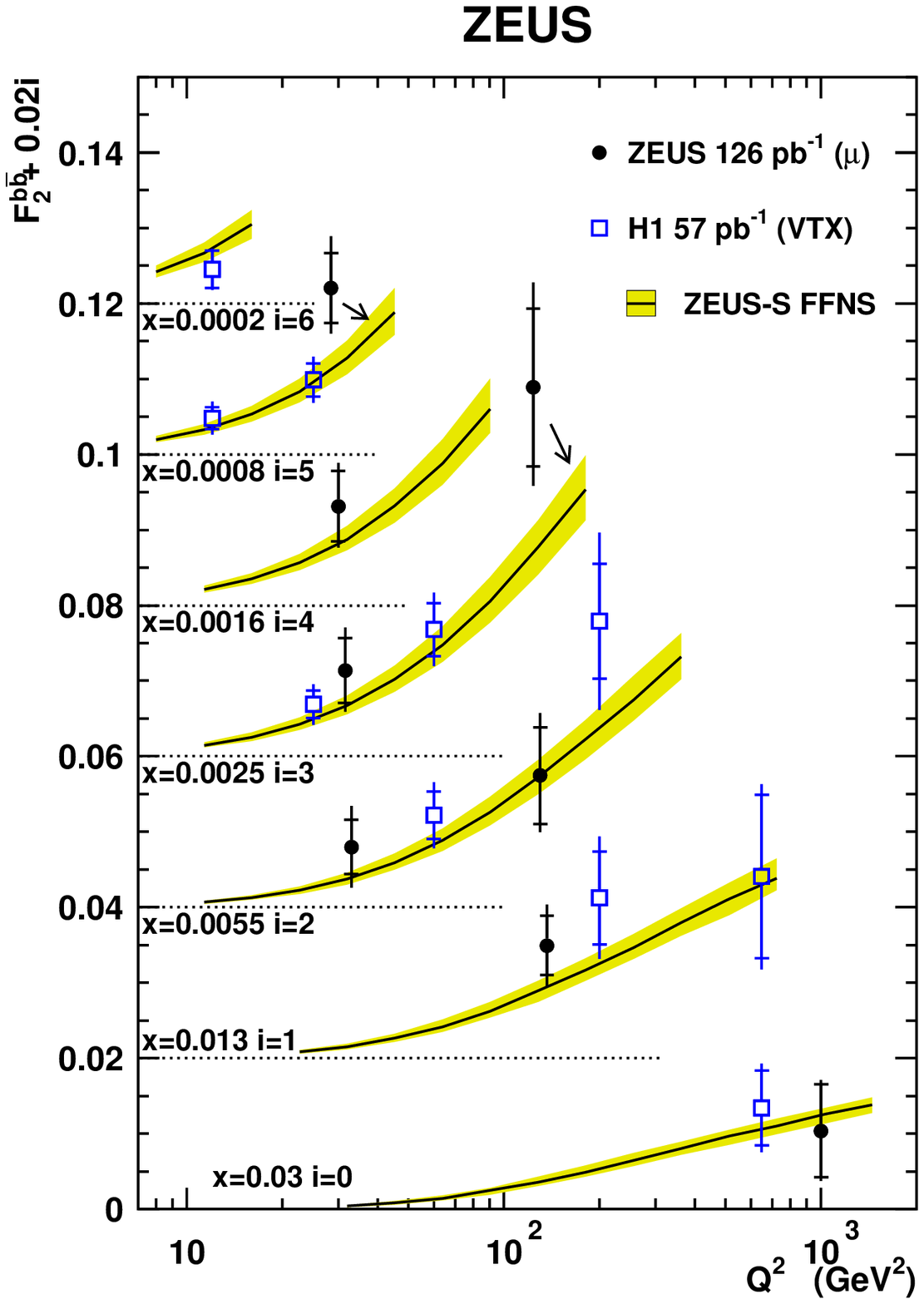,width=13.5 cm,clip=}
\end{center}
\caption{The structure function $F_2^{b\bar{b}}$ (filled symbols) plotted as a
function of $Q^2$ for fixed $x$ values.
The curves represent the  NLO QCD predictions 
in the FFNS using the ZEUS-S PDF fit.
The inner error bars are the statistical uncertainty while
the outer bars represent the statistical, systematic and theoretical uncertainties added in quadrature. The band represents the uncertainty on the NLO QCD
prediction. All the previous data (open symbols) are
also shown, corrected to the reference $x$ values used in this analysis:
$0.0005 \rightarrow 0.0008$, 
$0.002 \rightarrow 0.0025$,
$0.005 \rightarrow 0.0055$,
$0.032 \rightarrow 0.030$.
}
\label{f7}
\vfill
\end{figure}

The structure functions  $F_2^{c\bar{c}}$ and  $F_2^{b\bar{b}}$ are also presented
 in Figs.~\ref{f6} and \ref{f7} 
as  functions of $Q^2$ for fixed values of $x$, compared to
previous results corrected to the same reference $x$ used in the present
analysis.
%


\section{Summary}
\label{sec-concl}
The production of charm and beauty quarks was measured in DIS
using their decay into muons. Total and differential cross
sections for
muons from $c$ and $b$ decays  were measured in the kinematic region
$$Q^2>20\gev^2;\; 0.01<y<0.7;\; p_T^{\mu}>1.5\gev;\; -1.6<\eta^{\mu}<2.3$$
and compared to NLO QCD calculations. 
The agreement is good for charm.
Beauty is about a factor two above the central QCD prediction
although still compatible within statistical and systematic uncertainties.
The heavy quark contribution to the proton structure function $F_2$ was 
also measured and found to agree well with other measurements based
on independent techniques. For $Q^2\ge 60\gev^2$ 
the present results are of comparable or higher precision
than those previously existing.

\section{Acknowledgements}
 We appreciate the contributions to the construction and maintenance of the
 ZEUS detector of many people who are not listed as authors. The HERA
 machine group and the DESY computing staff are especially acknowledged for
 their success in providing excellent operation of the collider and the
 data analysis environment. We thank the DESY directorate for their strong
 support and encouragement. It is also a pleasure to thank Pedro Jimenez
 Delgado, Stefano Forte, Albero Guffanti,  Eric Laenen, Pavel Nadolsky, 
 Jack Smith and Paul Thompson for many illuminating discussions.
\vfill\eject

{
\def\bibname{\Large\bf References}
\def\refname{\Large\bf References}
\pagestyle{plain}
\ifzeusbst
  \bibliographystyle{./BiBTeX/bst/l4z_default}
\fi
\ifzdrftbst
  \bibliographystyle{./BiBTeX/bst/l4z_draft}
\fi
\ifzbstepj
  \bibliographystyle{./BiBTeX/bst/l4z_epj}
\fi
\ifzbstnp
  \bibliographystyle{./BiBTeX/bst/l4z_np}
\fi
\ifzbstpl
  \bibliographystyle{./BiBTeX/bst/l4z_pl}
\fi
{\raggedright
\bibliography{./BiBTeX/user/syn.bib,%
              ./BiBTeX/bib/l4z_articles.bib,%
              ./BiBTeX/bib/l4z_books.bib,%
              ./BiBTeX/bib/l4z_conferences.bib,%
              ./BiBTeX/bib/l4z_h1.bib,%
              ./BiBTeX/bib/l4z_misc.bib,%
              ./BiBTeX/bib/l4z_old.bib,%
              ./BiBTeX/bib/l4z_preprints.bib,%
              ./BiBTeX/bib/l4z_replaced.bib,%
              ./BiBTeX/bib/l4z_temporary.bib,%
              ./BiBTeX/bib/l4z_zeus.bib}}
}
\vfill\eject

\begin{table}[p]
\begin{center}
\begin{tabular}{|r@{ : }l|cll|cll|l|}
\hline
\multicolumn{2}{|c|}{$p_T^{\mu}$} &$d\sigma^{c}/dp_T^{\mu}$&$\Delta_{\rm stat.}$&$\Delta_{\rm syst.}$&
                                  $d\sigma^{b}/dp_T^{\mu}$&$\Delta_{\rm stat.}$&$\Delta_{\rm syst.}$&$\rho_{c,b}$ \\
\multicolumn{2}{|c|}{(GeV)}      &\multicolumn{3}{c|}{(pb/GeV)}
                                 &\multicolumn{3}{c|}{(pb/GeV)}&               \\ \hline
$1.5$&$ ~\,2.5  $&$  113    $&$\pm  10  $&$^{+21}_{-23 }       $&$   48    $&$\pm   8    $&$^{+15}_{-13} $&$   -0.48$\\
$2.5$&$ ~\,3.5  $&$  32.8   $&$\pm   3.8  $&$^{+ 5.7}_{ -5.4 } $&$   15.4  $&$ \pm  2.7   $&$^{+ 4.1}_{ -2.8}  $&$   -0.46$\\
$3.5$&$ ~\,5.0  $&$   6.0   $&$\pm   1.4  $&$^{+ 1.4}_{ -1.6 } $&$    6.3  $&$ \pm  1.1   $&$^{+ 0.9}_{-0.7} $&$   -0.48$\\
$5.0$&$ 10.0 $&$      0.97  $&$\pm   0.21 $&$^{+ 0.13}_{-0.20} $&$    0.76 $&$ \pm  0.22  $&$^{+ 0.10}_{-0.15} $&$   -0.45$\\
\hline\hline
\multicolumn{2}{|c|}{$\eta^{\mu}$} &$d\sigma^{c}/d\eta^{\mu}$&$\Delta_{\rm stat.}$&$\Delta_{\rm syst.}$&
                                  $d\sigma^{b}/d\eta^{\mu}$&$\Delta_{\rm stat.}$&$\Delta_{\rm syst.}$&$\rho_{c,b}$ \\
\multicolumn{2}{|c|}{}           &\multicolumn{3}{c|}{(pb)}
                                 &\multicolumn{3}{c|}{(pb)}&               \\ \hline
$ -1.60$&$-0.90 $ &$ 20.4 $&$\pm 3.3 $&$^{+ 4.2}_{ -4.1} $&$  5.0 $&$\pm 2.0 $&$^{+ 1.9}_{ -1.6} $&$ -0.36$\\
$ -0.90$&$-0.40 $ &$ 40.7 $&$\pm 6.0 $&$^{+ 7.7}_{ -8.0} $&$ 13.6 $&$\pm 3.5 $&$^{+ 4.1}_{ -1.9} $&$ -0.20$ \\
$ -0.40$&$+0.00 $ &$ 60.9 $&$\pm 7.8 $&$^{+11.8}_{-14.2} $&$ 17.3 $&$\pm 4.8 $&$^{+ 4.9}_{ -3.5} $&$ -0.37$\\
$ +0.00$&$+0.50 $ &$ 67.0 $&$\pm 7.1 $&$^{+10.2}_{-12.3} $&$ 21.2 $&$\pm 4.4 $&$^{+ 6.4}_{ -3.6} $&$ -0.42$\\
$ +0.50$&$+1.48 $ &$ 47.7 $&$\pm 6.4 $&$^{+ 8.9}_{ -8.8} $&$ 20.7 $&$\pm 4.3 $&$^{+ 6.0}_{ -5.5} $&$ -0.49$\\
$ +1.48$&$+2.30 $ &$ 33.4 $&$\pm 10.0 $&$^{+15.3}_{ -8.9} $&$16.4 $&$\pm 6.9 $&$^{+ 5.7}_{ -8.4} $&$ -0.41$\\ \hline
\hline
\multicolumn{2}{|c|}{$Q^2$} &$d\sigma^{c}/dQ^2$&$\Delta_{\rm stat.}$&$\Delta_{\rm syst.}$&
                             $d\sigma^{b}/dQ^2$&$\Delta_{\rm stat.}$&$\Delta_{\rm syst.}$&$\rho_{c,b}$ \\
\multicolumn{2}{|c|}{(GeV$^2$)}  &\multicolumn{3}{c|}{(pb/GeV$^2$)}
                                 &\multicolumn{3}{c|}{(pb/GeV$^2$)}&               \\ \hline
 $20$&$~~~\,\,\,40$&$  3.43   $&$\pm  0.40   $&$^{+0.72}_{   -0.66}   $&$  1.46    $&$\pm  0.24    $&$^{+0.30}_{   -0.39}  $&$-0.44$\\
 $40$&$~~~\,\,\,80$&$  1.22   $&$\pm  0.13   $&$^{+0.18}_{   -0.22}   $&$  0.546   $&$\pm  0.086   $&$^{+0.109}_{  -0.098} $&$-0.41$\\
 $80$&$~~\,\,200$&$  0.289  $&$\pm  0.031  $&$^{+0.053}_{  -0.054}  $&$  0.124   $&$\pm  0.023   $&$^{+0.020}_{  -0.019} $&$-0.36$\\
$200$&$~~\,\,500$&$  0.0447$ &$\pm  0.0071 $&$^{+0.0050}_{ -0.0083}$&$  0.0131$ &$\pm  0.0049 $&$^{+0.0035}_{ -0.0024}$&$-0.47$\\
$500$&$10000$&    $0.00063$  &$\pm  0.00013$&$^{+0.00014}_{-0.00010}$& $0.00013$  &$\pm  0.00008$&$^{+0.00005}_{-0.00003}$&$-0.38$\\
\hline\hline
\multicolumn{2}{|c|}{$x$} &$d\sigma^{c}/dx$&$\Delta_{\rm stat.}$&$\Delta_{\rm syst.}$&
                           $d\sigma^{b}/dx$&$\Delta_{\rm stat.}$&$\Delta_{\rm syst.}$&$\rho_{c,b}$ \\
\multicolumn{2}{|c|}{}           &\multicolumn{3}{c|}{(nb)}
                                 &\multicolumn{3}{c|}{(nb)}&               \\ \hline
$ 0.0003$&$  0.0010$&$ 35.3  $&$\pm  5.6 $&$^{+10.2}_{  -6.2  }$&$  17.4 $&$\pm  3.9  $&$^{+ 3.5}_{ -3.6}  $&$  -0.16$\\
$ 0.0010 $&$ 0.0020$&$ 35.2  $&$\pm  4.1 $&$^{+4.6}_{  -7.1  }$&$  12.4 $&$\pm  2.6  $&$^{+ 2.7}_{ -3.0}  $&$  -0.39$\\
$ 0.0020 $&$ 0.0040$&$ 16.1  $&$\pm  2.2 $&$^{+3.8}_{  -3.7  }$&$  \,~8.0 $&$\pm  1.4  $&$^{+ 1.6}_{ -1.2}  $&$  -0.51$\\
$ 0.0040 $&$ 0.0100$&$ \,~7.38 $&$\pm  0.72$&$^{+1.22}_{ -1.30 }$&$\,~  2.04 $&$\pm  0.45 $&$^{+ 0.52}_{-0.39} $&$  -0.45$\\
$ 0.0100 $&$ 0.1000$&$ \,~ 0.417$&$\pm 0.050$&$^{+0.068}_{-0.081}$&$\,~ 0.076 $&$\pm 0.028 $&$^{+0.030}_{-0.014}$&$ -0.53$\\
\hline
\end{tabular}
\caption{Muon differential cross sections for charm and beauty as a function of $\eta^{\mu}$,  $p_T^{\mu}$, $Q^2$ and $x$. The last column shows the statistical correlation coefficient between charm and beauty.}
  \label{t1}
\end{center}
\end{table}

\begin{table}[p]
\begin{center}
\begin{tabular}{|c|r@{ :}r|r@{ : }l|rll|rll|c|}
\hline
bin&\multicolumn{2}{c|}{$Q^2$}&    \multicolumn{2}{c|}{$x$} &$\sigma^{c}$& $\Delta_{\rm stat.}$&$\Delta_{\rm syst.}$&  
                                     $\sigma^{b}$&  $\Delta_{\rm stat.}$&$\Delta_{\rm syst.}$&$\rho_{c,b}$ \\
   &\multicolumn{2}{c|}{(GeV$^2$)}&\multicolumn{2}{c|}{}    &            &           (pb) &                       &
                                                                         & (pb)           &             &       \\ \hline
   1 &$ 20$&$  60$&$ 0.0003$&$ 0.0012$&$  32.9  $&$\pm 4.6  $&$^{+ 8.1}_{  -5.9} $&$  13.9 $&$\pm  2.9 $&$^{+  3.2}_{  -2.3}$&$ -0.29$\\
   2 &$ 20$&$  60$&$ 0.0012$&$ 0.0020$&$  17.7  $&$\pm 3.1  $&$^{+ 1.9}_{  -4.9}  $&$ 5.7 $&$\pm  2.0 $&$^{+  1.4}_{  -1.2}$&$ -0.42$\\
   3 &$ 20$&$  60$&$ 0.0020$&$ 0.0035$&$  16.2  $&$\pm 3.3  $&$^{+ 3.7}_{  -3.0}  $&$ 5.5 $&$\pm  2.0 $&$^{+  1.8}_{  -1.6}$&$ -0.51$\\
   4 &$ 20$&$  60$&$ 0.0035$&$ 0.0060$&$  35.1  $&$\pm 5.7  $&$^{+10.9}_{  -7.2}  $&$ 7.9 $&$\pm  3.6 $&$^{+  4.2}_{  -4.0}$&$ -0.56$\\
   5 &$ 60$&$ 400$&$ 0.0009$&$ 0.0035$&$  17.2  $&$\pm 2.7  $&$^{+ 3.8}_{  -2.7}  $&$ 8.8 $&$\pm  1.9 $&$^{+  1.6}_{  -1.4}$&$ -0.38$\\
   6 &$ 60$&$ 400$&$ 0.0035$&$ 0.0070$&$  18.4  $&$\pm 2.3  $&$^{+ 3.0}_{  -3.4}  $&$ 4.2 $&$\pm  1.5 $&$^{+  1.2}_{  -0.9}$&$ -0.35$\\
   7 &$ 60$&$ 400$&$ 0.0070$&$ 0.0400$&$  33.6  $&$\pm 3.5  $&$^{+ 6.1}_{  -6.4}  $&$ 8.6 $&$\pm  2.3 $&$^{+  2.1}_{  -2.1}$&$ -0.46$\\
   8 &$400$&$ 10000$&$0.0050$&$ 1.0000$&$   7.6  $&$\pm 1.5  $&$^{+ 1.2}_{  -1.4}  $&$ 1.6 $&$\pm  0.9 $&$^{+  0.4}_{  -0.4}$&$ -0.45$\\ \hline
\end{tabular}
\caption{Muon cross sections for charm and beauty in bins of $Q^2$ and $x$.
 The last column shows the statistical correlation coefficient between charm and beauty.}
  \label{t2}
\end{center}
\end{table}

\begin{table}
\begin{center}
\begin{tabular}{|c|r|c|c|c|c|c|c|c|}
\hline
bin & $Q^2$~(GeV$^2$) &$x$& $F_2^{c\bar{c}}$ & $\Delta_{\rm stat.}$ & $\Delta_{\rm syst.}$  &  $\Delta_{\rm theo.}$ & ${\cal A}$ & $C_L$ \\ \hline  
   1&    $30$& $0.0008$&  $0.318$ &$ \pm 0.044$ &$^{+0.078}_{-0.057} $&$^{+0.061}_{-0.042} $&$ 0.096 $&$ 0.980$ \\ 
   2&    $30$& $0.0016$&  $0.219$ &$ \pm 0.038$ &$^{+0.024}_{-0.061} $&$^{+0.043}_{-0.016} $&$ 0.114 $&$ 0.996$\\ 
   3&    $30$& $0.0025$&  $0.176$ &$ \pm 0.036$ &$^{+0.040}_{-0.033} $&$^{+0.032}_{-0.021} $&$ 0.113 $&$ 0.998$\\ 
   4&    $30$& $0.0055$&  $0.143$ &$ \pm 0.023$ &$^{+0.044}_{-0.029} $&$^{+0.028}_{-0.009} $&$ 0.096 $&$ 1.000$\\ 
   5&   $130$& $0.0025$&  $0.298$ & $ \pm0.047$ &$^{+0.066}_{-0.046} $&$^{+0.044}_{-0.025} $&$ 0.175 $&$ 0.955$\\ 
   6&   $130$& $0.0055$&  $0.228$ & $ \pm0.029$ &$^{+0.037}_{-0.042} $&$^{+0.030}_{-0.015} $&$ 0.220 $&$ 0.993$\\ 
   7&   $130$& $0.0130$&  $0.151$ & $ \pm0.016$ &$^{+0.027}_{-0.029} $&$^{+0.021}_{-0.011} $&$ 0.209 $&$ 0.999$\\ 
   8&  $1000$& $0.0300$&  $0.114$ & $ \pm0.023$ &$^{+0.018}_{-0.021} $&$^{+0.010}_{-0.007} $&$ 0.371 $&$ 0.987$\\ \hline
\end{tabular}
\caption{The structure function $F_2^{c\bar{c}}(x,Q^2)$. The last two columns show the muon kinematic acceptance, ${\cal A}$, and the correction for the longitudinal structure function, $C_L$.}
  \label{t3a}
\end{center}
\end{table}

\begin{table}
\begin{center}
\begin{tabular}{|c|r|c|c|c|c|c|c|c|}
\hline
bin & $Q^2$(GeV$^2$)   &$x$& $F_2^{b\bar{b}}$ & $\Delta_{\rm stat.}$ & $\Delta_{\rm syst.}$  &  $\Delta_{\rm theo.}$ & ${\cal A}$ & $C_L$ \\ \hline   
1&       $30 $&$ 0.0008 $&$ 0.0220 $&$\pm 0.0047 $&$^{+0.0049}_{-0.0037} $&$ ^{+0.0011}_{-0.0010} $&$ 0.260 $&$ 0.992$\\
2&       $30 $&$ 0.0016 $&$ 0.0131 $&$\pm 0.0047 $&$^{+0.0032}_{-0.0028} $&$ ^{+0.0009}_{-0.0003} $&$ 0.264 $&$ 0.998$\\
3&       $30 $&$ 0.0025 $&$ 0.0114 $&$\pm 0.0043 $&$^{+0.0037}_{-0.0034} $&$ ^{+0.0005}_{-0.0004} $&$ 0.251 $&$ 0.999$\\
4&       $30 $&$ 0.0055 $&$ 0.0080 $&$\pm 0.0036 $&$^{+0.0041}_{-0.0041} $&$ ^{+0.0004}_{-0.0003} $&$ 0.189 $&$ 1.000$\\
5&      $130 $&$ 0.0025 $&$ 0.0489 $&$\pm 0.0105 $&$^{+0.0088}_{-0.0076} $&$ ^{+0.0024}_{-0.0018} $&$ 0.300 $&$ 0.962$\\
6&      $130 $&$ 0.0055 $&$ 0.0175 $&$\pm 0.0064 $&$^{+0.0052}_{-0.0039} $&$ ^{+0.0007}_{-0.0007} $&$ 0.319 $&$ 0.994$\\
7&      $130 $&$ 0.0130 $&$ 0.0149 $&$\pm 0.0039 $&$^{+0.0037}_{-0.0037} $&$ ^{+0.0007}_{-0.0006} $&$ 0.281 $&$ 0.999$\\
8&     $1000 $&$ 0.0300 $&$ 0.0104 $&$\pm 0.0061 $&$^{+0.0028}_{-0.0025} $&$ ^{+0.0004}_{-0.0004} $&$ 0.420 $&$ 0.983$\\
\hline
\end{tabular}
\caption{The structure function $F_2^{b\bar{b}}(x,Q^2)$. The last two columns show the muon kinematic acceptance, ${\cal A}$, and the correction for the longitudinal structure function, $C_L$.}
  \label{t3b}
\end{center}
\end{table}


\vfill

\appendix

\section*{Appendix:\\ Tables of systematic and theoretical uncertainties}


\begin{table}[h!]
\begin{center}
{\footnotesize
\begin{tabular}{|r|rrrrrrrr|rrrrrrrr|}
\hline
Syst.  &\multicolumn{8}{c|}{$F_2^{c\bar{c}}$ bin}&\multicolumn{8}{c|}{$F_2^{b\bar{b}}$ bin}\\ 

   & 1 & 2 & 3 & 4 & 5 & 6 & 7 & 8 & 1 & 2 & 3 & 4 & 5 & 6& 7& 8 \\ \hline\hline
 1a& -5& -6& -6& -4& -6& -6& -6& -6& -6& -5& -2&  7& -6&-6& -2& -5\\
 1b&  6&  7&  6&  4&  6&  6&  6&  7&  7&  5&  1& -7&  6& 6&  1&  4\\ \hline
 2a&  0&  1& -1& -6&  1&  0& -1&  1& -2& -6&-12&-37& -2&-4&-14& -6\\
 2b&  0& -1&  1&  5& -1&  0&  1& -1&  2&  5& 11& 43&  2& 3& 14&  6\\ \hline
 3a& -3& -2& -3& -3& -4& -3& -2& -2&  0& -2&  0& -8&  1& 1& -2& -3\\
 3b&  3&  2&  3&  4&  4&  3&  2&  1&  0&  3&  1& 11& -1&-1&  4&  5\\ \hline
 4a&  2&-11&  8& -3& -1& -6& -6& -9&  2&  7& -8& 14& -7&10&  1& -8\\
 4b& -2& -1&  2&  5& 15&  2&  5& 12&  9& -9& -8& 13& -7&11&  0& -4\\ \hline
 5a&  4&  4& 17& 20&  9& 11& 12&  3&  2& -3&  0&  5& -2& 4& -6&  3\\
 5b& -4&-20&-10&-15& -1& -9&-11& -2&  5& 11& -1& 10& -3&-9&  3& -5\\ \hline
 6a& 19& -5&  1&  1&  0& -2&-10&  0&  3&  5&-13&-28& -2&-9& -3& -7\\
 6b& -9&  0&  3& 19&  4&  2&  8&  3&  6& -3& 22&  3&  2&11& -1&  4\\ \hline
  7&  8& -5& -2&  1& -5& -4&  4&  4&  1& -8&  7& -8& -4&11&-16& -9\\ \hline
 8a&  0& -2& -1&  1&  0&  2& -1& -3& 11&  6& -4& -1&  8& 6& 11& 21\\
 8b&  0&  4& -3&  3&  3&  0& -3&  4&-11&-10& -5&-10& -5&-7& -1&-16\\ \hline
 9a&  2& -1& -4&  1&  0& -1& -2& -1& -5&  7& 14&  2&  5&10&  8&  6\\
 9b& -1&  1&  3& -1&  0&  1&  1&  0&  4& -4& -9& -1& -3&-6& -5& -3\\ \hline
10a& -2&  4&  1&  1& -2& -1&  0&  0& -1& -4& -1& -1& -1&-1&  2&  1\\
10b&  1& -5&  0&  2& -2& -1&  2& -1& -4&  3&  4& -5&  6&-6&  0&  5\\ \hline
11a&  0& -1&  3&  9&  6& -2&  1&  0& -5&  6& 11& -8& -3&-2& -1&  3\\
11b&  0& -7& -9& -4&  2&  2&  0&  0&  1&  5&  3&  6& -2&-3&  1&  2\\ \hline
12a& -3&  0& -2&  0& -3& -3& -2&  0&  2& -2&  2& -7&  4& 6&  4&  5\\
12b&  2&  0&  1& -1&  2&  2&  2& -1& -1&  2& -1&  9& -3&-5& -3& -3\\ \hline
 13& -1& -1& -2&  1& -2&  0& -1&  0&  9& 10&  5& -2&  9& 5&  6&  1\\ \hline
14a&  6&  1&  3&  1&  4&  6&  3&  0&  4&  5&  5&  3& -2&-8&  6&  8\\
14b&-12& -7& -8& -9&-11&-12& -7&-10& -6& -8&-11&  4&  3&12& -6&  2\\ \hline
15a& -3& -3& -3& -2& -3& -3& -3& -3& -3& -2& -1&  3& -3&-2& -1& -2\\
15b&  3&  3&  3&  2&  3&  3&  3&  3&  3&  2&  1& -3&  3& 2&  1&  2\\ \hline
16 &  6& -2&  5& -5&  4&  3& -2& -8&  1&  6&-12& -5&  0&-6&  5&  2\\ \hline
17a& -3& -3& -3& -3& -3& -3& -3& -3& -3& -3& -3& -3& -3&-3& -3& -3\\
17b&  3&  3&  3&  3&  3&  3&  3&  3&  3&  3&  3&  3&  3& 3&  3&  3\\ \hline
\end{tabular}
}
\caption{Systematic uncertainties of the $F_2^{c\bar{c}}$ and  $F_2^{c\bar{c}}$ measurements. The first column gives the systematic variation number as reported in Section 8, with ``a'' and ``b'' corresponding to variations in opposite directions. The other columns list the effect of each variation on the measurements in percent.}
  \label{t4}
\end{center}
\end{table}

\begin{table}
\begin{center}
{\footnotesize
\begin{tabular}{|r|rrrrrrrr|rrrrrrrr|}
\hline
Syst.  &\multicolumn{8}{c|}{$F_2^{c\bar{c}}$ bin}&\multicolumn{8}{c|}{$F_2^{b\bar{b}}$ bin}\\ 

                & 1   & 2  & 3  & 4  & 5  &  6 &  7 &  8 &  1 &  2 &  3 &  4 &  5 &  6 &  7 & 8 \\ \hline\hline
$\mu_F \times2$ &$  4$&$ 3$&$ 1$&$ 2$&$ 1$&$ 1$&$ 0$&$ 0$&$ 1$&$ 2$&$ 0$&$-1$&$ 0$&$ 0$&$-1$&$ 0$\\
$/2$            &$ -8$&$-2$&$-7$&$ 0$&$-4$&$-2$&$ 0$&$ 1$&$-1$&$ 0$&$-1$&$ 0$&$-1$&$ 0$&$ 1$&$ 0$\\ \hline
$\mu_R \times2$ &$  1$&$-1$&$-3$&$-2$&$-2$&$-2$&$-3$&$-2$&$ 0$&$ 1$&$ 0$&$-2$&$ 0$&$ 0$&$-1$&$-1$\\
$/2$            &$ -1$&$ 0$&$ 2$&$ 6$&$ 1$&$ 2$&$ 5$&$ 3$&$-1$&$ 1$&$ 1$&$ 1$&$ 1$&$ 0$&$ 1$&$ 1$\\ \hline
$m_q +$         &$ -8$&$-5$&$-3$&$-2$&$-3$&$-1$&$-1$&$-2$&$-2$&$-1$&$-1$&$ 0$&$-1$&$-1$&$ 0$&$-1$\\
    $-$         &$  5$&$ 4$&$ 3$&$ 3$&$ 3$&$ 2$&$ 2$&$ 1$&$ 2$&$ 2$&$ 1$&$ 0$&$ 1$&$ 1$&$ 2$&$ 0$\\ \hline
PDF $+$         &$  2$&$-1$&$-1$&$ 0$&$ 1$&$ 0$&$ 2$&$ 0$&$ 0$&$ 1$&$-1$&$-1$&$ 0$&$ 0$&$-1$&$-1$\\
    $-$         &$ -1$&$ 1$&$-2$&$-1$&$-1$&$ 0$&$ 0$&$ 0$&$ 0$&$ 1$&$ 0$&$ 1$&$ 0$&$ 0$&$ 1$&$ 0$\\ \hline
CTEQ            &$  4$&$ 1$&$-1$&$ 1$&$ 1$&$ 0$&$ 2$&$ 2$&$ 0$&$ 1$&$-1$&$ 3$&$ 0$&$ 1$&$ 2$&$1$\\ \hline
$\epsilon +$    &$ 16$&$17$&$17$&$17$&$13$&$12$&$12$&$ 7$&$ 2$&$ 4$&$ 2$&$ 2$&$ 3$&$ 3$&$ 3$&$ 2$\\
         $-$    &$ -6$&$-2$&$-8$&$-4$&$-5$&$-4$&$-5$&$-4$&$-3$&$-1$&$-3$&$-2$&$-3$&$-3$&$-3$&$-2$\\ \hline
$E+p_{||}$           &$  5$&$ 6$&$ 4$&$ 4$&$ 3$&$ 3$&$ 2$&$ 0$&$ 3$&$ 5$&$ 3$&$ 2$&$ 3$&$ 3$&$ 2$&$ 0$\\ \hline
${\cal B} +$    &$ -4$&$-4$&$-4$&$-4$&$-4$&$-4$&$-4$&$-4$&$-2$&$-2$&$-2$&$-2$&$-2$&$-2$&$-2$&$-2$\\
         $-$    &$  4$&$ 4$&$ 4$&$ 4$&$ 4$&$ 4$&$ 4$&$ 4$&$ 2$&$ 2$&$ 2$&$ 2$&$ 2$&$ 2$&$ 2$&$ 2$\\ \hline
\end{tabular}
}
\caption{Theoretical uncertainties of the $F_2^{c\bar{c}}$ and  $F_2^{c\bar{c}}$ measurements. The first column gives the parameter varied in the calculation as reported in Sections 2 and 10: the factorisation ($\mu_F$) and renormalisation ($\mu_R$) scales, the HQ mass ($m_q$), the variation of the ZEUS PDF by its uncertainty (PDF), the use of the CTEQ5 PDF (CTEQ), the Peterson fragmentation parameter ($\epsilon$), the use an alternative fragmentation variable ($E+p_{||}$) and the SL branching ratio (${\cal B}$).
The other columns list the effect of each variation on the measurements in percent.
}  \label{t5}
\end{center}
\end{table}

%
%
\end{document}